\documentclass[english,onecolumn,draftcls]{IEEEtran}
\usepackage[T1]{fontenc}
\usepackage[latin9]{inputenc}
\usepackage{verbatim}
\usepackage{amsthm}
\usepackage{amsmath}
\usepackage{amssymb}
\usepackage{graphicx}
\usepackage{esint}

\makeatletter
  \theoremstyle{definition}
  \newtheorem{example}{\protect\examplename}
  \theoremstyle{remark}
  \newtheorem{rem}{\protect\remarkname}
  \theoremstyle{definition}
  \newtheorem{defn}{\protect\definitionname}
  \theoremstyle{plain}
  \newtheorem{prop}{\protect\propositionname}
  \theoremstyle{plain}
  \newtheorem{lem}{\protect\lemmaname}
  \theoremstyle{plain}
  \newtheorem{thm}{\protect\theoremname}
  \theoremstyle{plain}
  \newtheorem{cor}{\protect\corollaryname}

\usepackage{cite}

\newtheorem{assumption}{Assumption}

\makeatother

\usepackage{babel}
\providecommand{\corollaryname}{Corollary}
\providecommand{\definitionname}{Definition}
\providecommand{\examplename}{Example}
\providecommand{\lemmaname}{Lemma}
\providecommand{\propositionname}{Proposition}
\providecommand{\remarkname}{Remark}
\providecommand{\theoremname}{Theorem}

\begin{document}

\title{Joint Power and Antenna Selection Optimization in Large Cloud Radio
Access Networks}

\author{{\normalsize{An Liu, }}\textit{\normalsize{Member IEEE}}{\normalsize{,
and Vincent Lau,}}\textit{\normalsize{ Fellow IEEE}}{\normalsize{,\\Department
of Electronic and Computer Engineering, Hong Kong University of Science
and Technology}}}
\maketitle
\begin{abstract}
Large multiple-input multiple-output (MIMO) networks promise high
energy efficiency, i.e., much less power is required to achieve the
same capacity compared to the conventional MIMO networks if perfect
channel state information (CSI) is available at the transmitter. However,
in such networks, huge overhead is required to obtain full CSI especially
for Frequency-Division Duplex (FDD) systems. To reduce overhead, we
propose a downlink antenna selection scheme, which selects $S$ antennas
from $M>S$ transmit antennas based on the large scale fading to serve
$K\leq S$ users in large distributed MIMO networks employing regularized
zero-forcing (RZF) precoding. In particular, we study the joint optimization
of antenna selection, regularization factor, and power allocation
to maximize the average weighted sum-rate. This is a mixed combinatorial
and non-convex problem whose objective and constraints have no closed-form
expressions. We apply random matrix theory to derive asymptotically
accurate expressions for the objective and constraints. As such, the
joint optimization problem is decomposed into subproblems, each of
which is solved by an efficient algorithm. In addition, we derive
structural solutions for some special cases and show that the capacity
of very large distributed MIMO networks scales as $O\left(K\textrm{log}M\right)$
when $M\rightarrow\infty$ with $K,S$ fixed. Simulations show that
the proposed scheme achieves significant performance gain over various
baselines.\end{abstract}

\section{Introduction}

Large MIMO networks have been a hot research topic due to their high
energy efficiency \cite{Marzetta_SPM12_LargeMIMO}. Such networks
are equipped with an order of magnitude more antennas than conventional
systems, i.e., a hundred antennas or more. In centralized large MIMO
systems where all antennas are collocated at the base station (BS),
high energy efficiency is realized by exploiting the increased spatial
degrees of freedom and beamforming gain. In large distributed MIMO
systems where the antennas are distributed geographically, enhanced
energy efficiency is achieved from shortened distances between antennas
and users as well as improved spectral efficiency per unit area. There
are a number of prior works on large MIMO networks, including various
topics such as information theoretical capacity \cite{Moustakas_TIT03_LASPC},
transceiver design \cite{Liang_TSP06_LASdete}, CSI acquisition, and
pilot contamination \cite{Marzetta_TWC11_PilotContamination}. In
particular, various downlink precoding schemes have been proposed
and analyzed. Remarkably, the simple zero-forcing (ZF) precoding is
shown to achieve most of the capacity of large MIMO downlink \cite{Marzetta_SPM12_LargeMIMO}.
One of the main challenges towards achieving the performance predicted
by the theoretical analysis is how to obtain the CSI at the transmitter
(CSIT) for a large number of antennas. In most of the existing works,
Time-Division Duplex (TDD) is assumed and channel reciprocity can
be exploited to obtain CSIT via uplink pilot training. In \cite{Muharar_ICC11_RCIRMT,Wagner_TIT12s_LargeMIMO},
random matrix theory is used to analyze the asymptotic performance
of ZF and RZF \cite{Peel_TOC05_RCI} in both TDD and FDD systems,
with a focus on the case when all the antennas are collocated at a
BS. For FDD systems, the amount of CSI feedback required to maintain
a constant per-user rate gap from the perfect CSIT case has also been
analyzed in \cite{Wagner_TIT12s_LargeMIMO} under the assumption of
perfect CSI estimation at the users. In practice, we need $M$ orthogonal
pilot sequences to estimate the channel corresponding to the $M$
transmit antennas. However, the number of available orthogonal pilot
sequences is limited by the channel coherent time and it may become
smaller than $M$ as $M$ grows large.

\begin{figure}
\begin{centering}
\textsf{\includegraphics[clip,width=85mm]{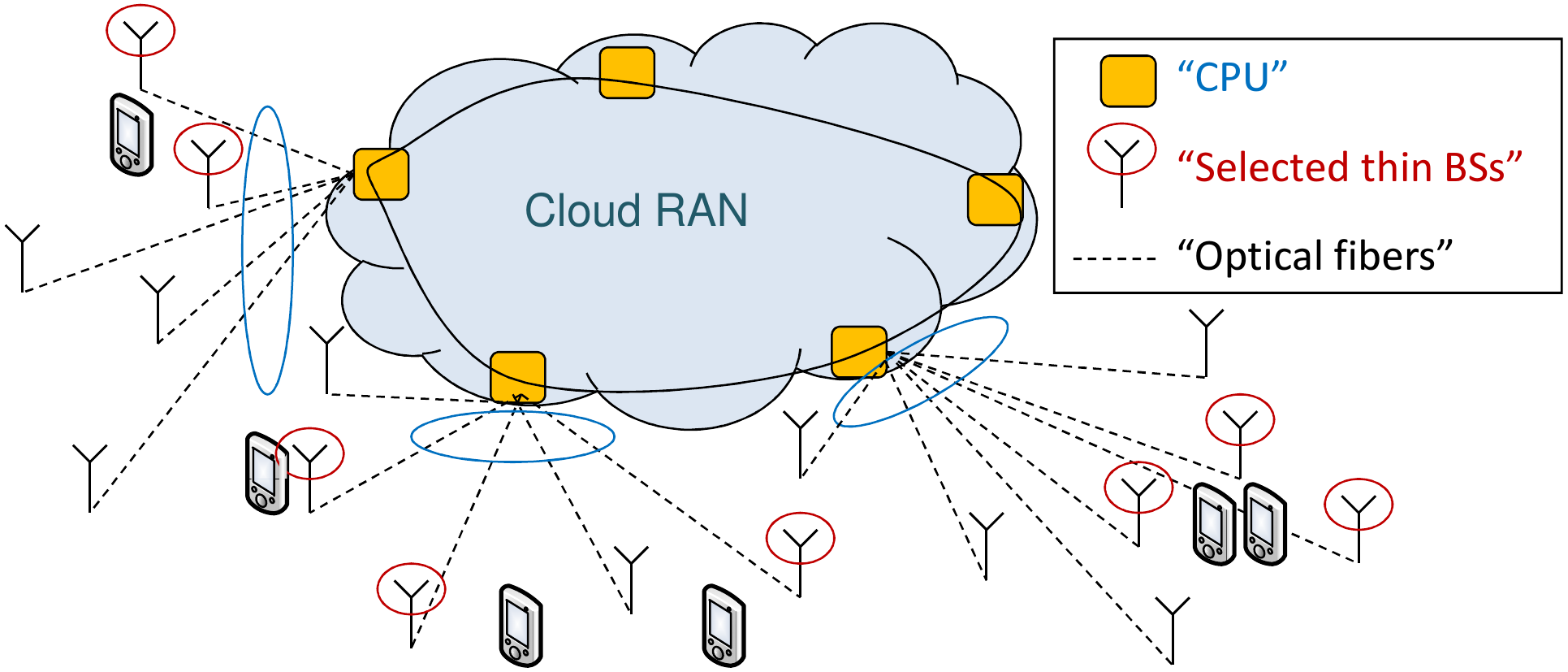}}
\par\end{centering}

\caption{\label{fig:system_model}Illustration of a large distributed MIMO
network, which consists of $M$ thin base stations (distributed antennas)
connected to a C-RAN via high speed optical fiber. }
\end{figure}

In this paper, we consider large distributed MIMO networks operating
in FDD mode in which there are $M$ distributed antennas (thin BSs%
\footnote{A thin base-station refers to a low cost and low power base station
and this name is borrowed from the nomenclature \textquotedbl{}thin
client\textquotedbl{} in cloud computing.%
}) linked together by high speed fiber backhaul as illustrated in Fig.
\ref{fig:system_model}. Such networks are also called the cloud radio
access networks (C-RAN) \cite{Huawei_CJKW11_cloudRan}. In such a
scenario, only a few nearby antennas can contribute significantly
to a user's communication due to path loss. To avoid expensive CSI
acquisition and signal processing overheads for antennas with huge
path losses to the users, a subset of $S$ antennas is selected to
serve a given set of $K$ users using RZF precoding \cite{Peel_TOC05_RCI},
where $M\gg S\geq K$. RZF precoding has been shown in \cite{Hanly_IT11s_Multicell_RMT}
to be asymptotically optimal for $S,K\rightarrow\infty$ in a two-cell
system. 

The existing antenna selection schemes in multi-user MIMO systems
\cite{Heath_GC07_ASBD,Chang_PIMRC09_ASDPC} require global knowledge
of the instantaneous CSI which is unacceptable for large $M$. In
3G and LTE systems, users are associated with the strongest antennas/BSs.
However, this baseline algorithm is inefficient when the antennas/BSs
are allowed to perform cooperative MIMO (CoMP) \cite{Irmer_Comm11_CoMPsurvey}
as illustrated in the following two examples. In both examples, we
assume $S=2$ distributed antennas are selected to serve $K=2$ users.

\begin{figure}
\centering{}%
\begin{minipage}[t]{0.48\textwidth}%
\includegraphics[clip,width=78mm]{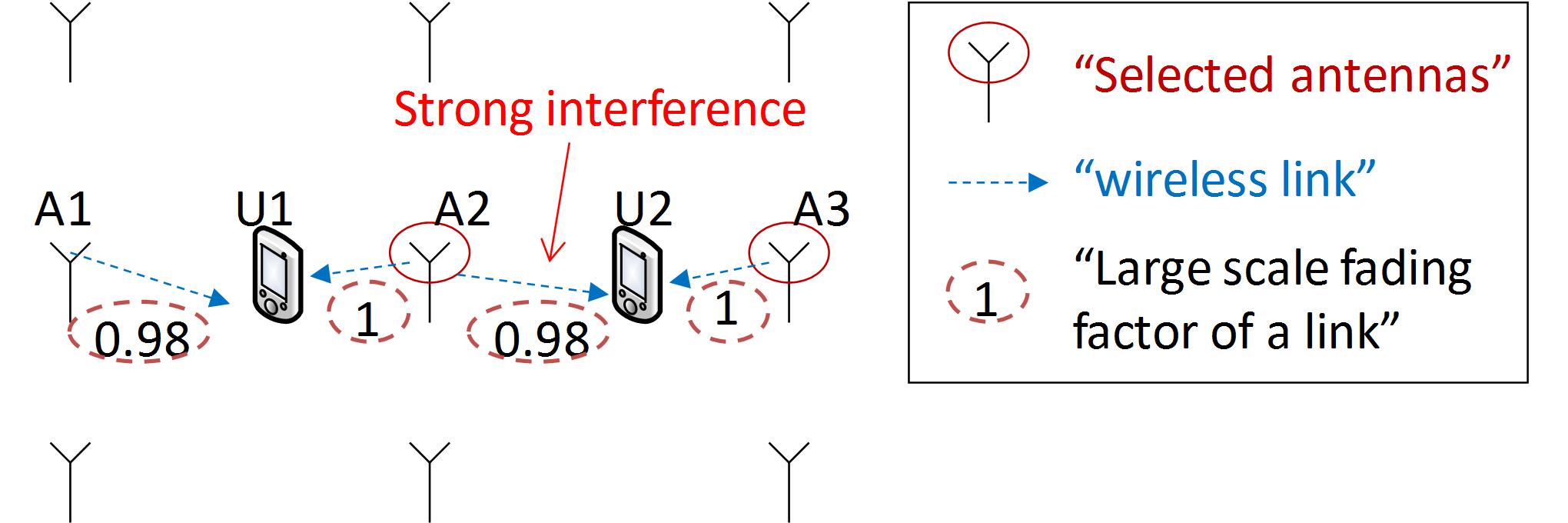}

\caption{\label{fig:example_SIFC}An example that strong cross link causes
large interference}
\end{minipage}\hfill{}%
\begin{minipage}[t]{0.48\textwidth}%
\includegraphics[clip,width=78mm]{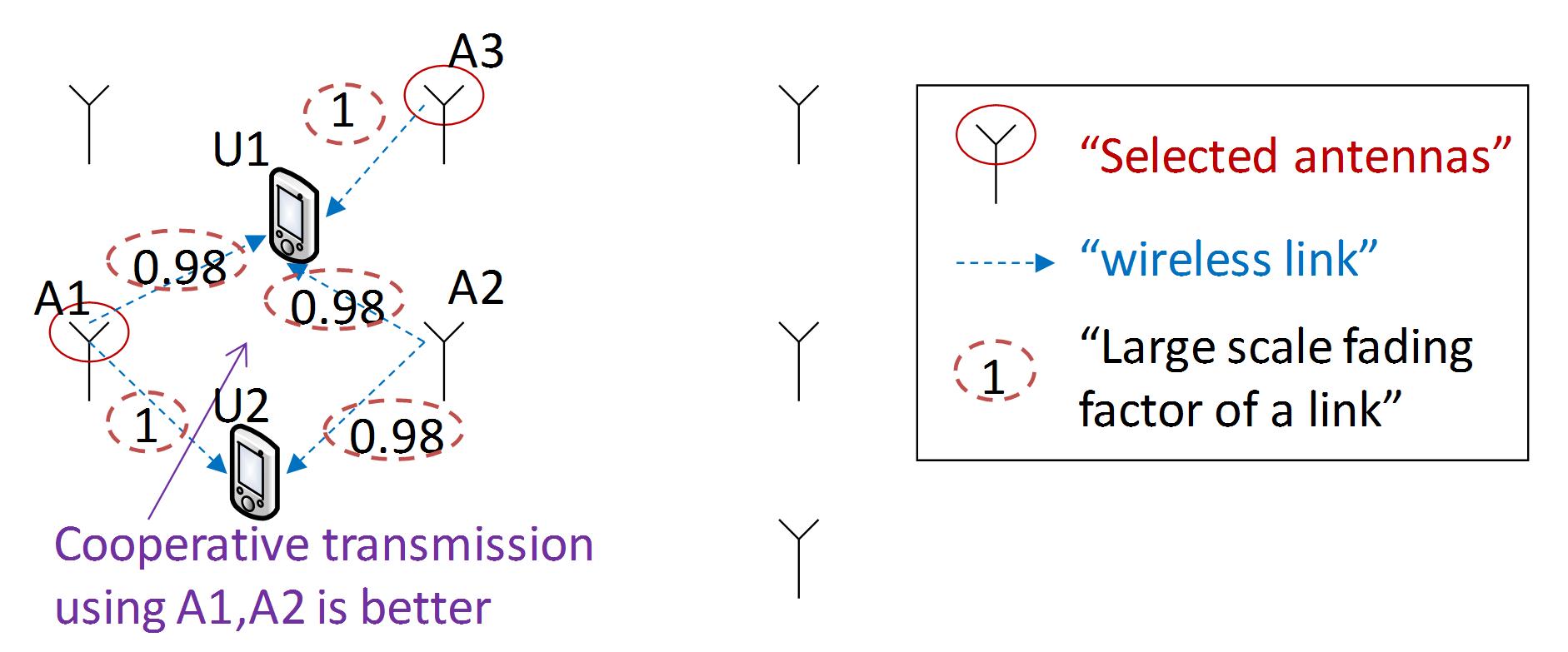}

\caption{\label{fig:example_Cop}An example that strong cross links provide
cooperative gain}
\end{minipage}
\end{figure}

\begin{example}
[Strong cross link causing low SINR]\label{exa:Strong-cross-link-causes}
Fig. \ref{fig:example_SIFC} illustrates the path loss configuration.
According to the baseline algorithm, the selected antennas will be
$\mathcal{A}=\left\{ 2,3\right\} $. However, this is not a good choice
because antenna 2 causes strong interference to user 2 before precoding.
Although the interference can be suppressed using RZF precoding, the
overall SINR is still low because the cross link from A3 to U1 is
weak and the joint transmission gain is limited. A better choice would
be $\mathcal{A}=\left\{ 1,3\right\} $.
\end{example}

\begin{example}
[Strong cross link providing cooperative gain]\label{exa:Strong-cross-link-provides}
Fig. \ref{fig:example_Cop} illustrates the path loss configuration.
According to the baseline algorithm, the selected antennas will be
$\mathcal{A}=\left\{ 1,3\right\} $. Instead, better performance can
be achieved by letting $\mathcal{A}=\left\{ 1,2\right\} $ due to
cooperative transmission.
\end{example}

Hence, a more efficient antenna selection design is crucial for C-RAN.
We study the joint optimization of antenna selection, regularization
factor in RZF precoding, and power allocation, to maximize the average
weighted sum-rate under per antenna power constraints. The optimization
only requires the knowledge of large scale fading factors and the
overhead for CSI acquisition is greatly reduced as discussed in Remark
\ref{rem:Comments-on-CSI}. The following are two first-order challenges.
\begin{itemize}
\item \textbf{Combinatorial Optimization Problem}: The antenna selection
problem with CoMP processing in the C-RAN is combinatorial with exponential
complexity w.r.t. the total number of antennas $M$.
\item \textbf{Asymptotic Performance Analysis}: It is important to derive
closed-form performance expressions in order to obtain design insights.
Yet, the performance analysis is non-trivial due to the heterogeneous
path loss as well as the lack of closed form antenna selection solution. 
\end{itemize}

In this paper, we extend the results in \cite{Wagner_TIT12s_LargeMIMO}
to obtain deterministic equivalent (DE) of the weighted sum-rate and
the per-antenna transmit power%
\footnote{We also noticed that the downlink channel $\mathbf{H}$ in C-RAN can
be modeled as the gram random matrices with a given variance profile
\cite{Hachem_Applprob09_CLTRMT}. The mutual information $\textrm{log}\left|\mathbf{H}\mathbf{H}^{\dagger}+\rho\mathbf{I}\right|$
for such channel model, where $\rho>0$ is a constant, has been shown
in \cite{Hachem_Applprob09_CLTRMT} to have a Gaussian limit whose
parameters are identified as the dimension of $\mathbf{H}$ goes to
infinity. In this paper, we focus on a different problem, i.e., the
joint optimization of power and antenna selection to maximize the
weighted sum-rate under RZF precoding.%
}. By exploiting the implicit structure in the objective and constraints
functions, the joint optimization problem is decomposed into simpler
subproblems, each of which is solved by an efficient algorithm. We
also show that there is an asymptotic decoupling effect in very large
distributed MIMO networks and the capacity grows logarithmically with
the total number of antennas $M$ even when the number of active antennas
$S$ is fixed.

The rest of the paper is organized as follows. The system model is
outlined in Section \ref{sec:System-Model}. In Section \ref{sec:Optimization-Formulation-for},
the antenna selection problem is formulated and its deterministic
approximation is derived using random matrix theory. The solution
of the problem is presented in Section \ref{sec:Optimization-Solution}.
In Section \ref{sec:Simplified-Solutions-and}, we give structural
solutions for some special cases. Simulations are used to verify the
performance of the proposed solution in Section \ref{sec:Numerical-Results}
and the conclusion is given in Section \ref{sec:Conlusion}.

\section{System Model\label{sec:System-Model}}

Consider the downlink of C-RAN with $M$ distributed transmit antennas
and $K$ single-antenna users as illustrated in Fig. \ref{fig:system_model}.
The $M\gg K$ distributed antennas are connected to a C-RAN \cite{Huawei_CJKW11_cloudRan}
via fiber backhaul and the system operates in FDD mode. Denote $h_{km}$
as the channel between the $m^{\textrm{th}}$ transmit antenna and
the $k^{\textrm{th}}$ user. We consider a composite fading channel,
i.e., $h_{km}=\sigma_{km}W_{km},\:\forall k,m,$ where $\sigma_{km}\geq0$
is the large scale fading factor caused by, e.g., path loss and shadow
fading, and $W_{km}$ is the small scale fading factor. 

\begin{assumption}[Channel model]\label{asm:forY}The small scale
fading process $W_{km}\left(t\right)\sim\mathcal{CN}\left(\mathbf{0},1\right)$
is quasi-static within a time slot but i.i.d. w.r.t. time slots and
the spatial indices $k,m$. The large scale fading process $\sigma_{km}\left(t\right)$
is assumed to be a slow ergodic random process (i.e., $\sigma_{km}\left(t\right)$
remains constant for a large number of time slots) according to a
general distribution.

\end{assumption}

The baseband processing is centralized at the C-RAN. To limit the
signaling overheads, we consider antenna selection where a subset
$\mathcal{A},\:\left|\mathcal{A}\right|=S\geq K$ of the $M$ antennas
are selected to serve the $K$ users. Let $\mathcal{A}_{j}$ denote
the $j^{\textrm{th}}$ element in $\mathcal{A}$. Let $\mathbf{H}\left(\mathcal{A}\right)\in\mathbb{C}^{K\times S}$
denote the composite downlink channel matrix between the selected
$S$ antennas and the $K$ users, and define $\mathbf{\Sigma}\left(\mathcal{A}\right)\in\mathbb{R}_{+}^{K\times S}$
as the corresponding large scale fading matrix, whose element at the
$k^{\textrm{th}}$ row and the $j^{\textrm{th}}$ column is $\sigma_{k\mathcal{A}_{j}}$.
For conciseness, $\mathbf{H}\left(\mathcal{A}\right)$ and $\mathbf{\Sigma}\left(\mathcal{A}\right)$
are denoted as $\mathbf{H}$ and $\mathbf{\Sigma}$ when there is
no ambiguity. 

\begin{assumption}[CSIT assumption]\label{asm:forY}The C-RAN has
knowledge of all the $K\times M$ large scale fading factors $\sigma_{km}$'s
and the $K\times S$ instantaneous channel matrix $\mathbf{H}\left(\mathcal{A}\right)$
corresponding to the selected antennas in $\mathcal{A}$ only.\hfill \IEEEQED

\end{assumption}
\begin{rem}
[CSI Acquisition]\label{rem:Comments-on-CSI}In FDD systems, $\mathbf{H}\left(\mathcal{A}\right)$
can be obtained via downlink channel estimation and channel feedback.
The amount of training for $\mathbf{H}\left(\mathcal{A}\right)$ is
limited by the channel coherence time, which depends on the user movement
speed. Hence, for large $M$, the estimated CSI quality at the C-RAN
will be poor if all the $M$ antennas in the network are active. Using
antenna selection and with properly chosen $S$, the instantaneous
CSI for the $S$ selected antennas can be estimated and fed back to
the C-RAN using conventional arrangement in LTE. Hence, the problem
of CSI limitation in C-RAN can be alleviated by antenna selection
based on large scale fading factors. On the other hand, the large
scale fading matrix $\mathbf{\Sigma}$ is a long-term statistic and
can be estimated at the C-RAN from the uplink reference signals \cite{Maretta_TSP01_UDSPCduality}
due to the reciprocity of large scale fading factors.\hfill \IEEEQED
\end{rem}

We consider the RZF precoding scheme \cite{Peel_TOC05_RCI}. The composite
receive signal vector for the $K$ users can be expressed as:
\[
\mathbf{y}=\mathbf{H}\mathbf{F}\mathbf{s}+\mathbf{n},
\]
where $\mathbf{s}=\left[s_{1},...,s_{K}\right]\sim\mathcal{CN}\left(\mathbf{0},\mathbf{I}_{K}\right)$
is the symbol vector; $\mathbf{n}\sim\mathcal{CN}\left(\mathbf{0},\mathbf{I}_{K}\right)$
is the AWGN noise vector; and $\mathbf{F}=\left[\mathbf{f}_{1},...,\mathbf{f}_{K}\right]\in\mathbb{C}^{S\times K}$
is the RZF precoding matrix given by
\begin{eqnarray}
\mathbf{F} & = & \left(\mathbf{H}^{\dagger}\mathbf{H}+\alpha S\mathbf{I}_{S}\right)^{-1}\mathbf{H}^{\dagger}\mathbf{P}^{1/2},\label{eq:PrecodingM}
\end{eqnarray}
where $\alpha$ is the regularization factor and $\mathbf{P}=\textrm{diag}\left(p_{1},...,p_{K}\right)$
is a power allocation matrix. Note that the regularization factor
is scaled by $S$ to ensure that $\alpha$ is bounded as $S,K\rightarrow\infty$
\cite{Wagner_TIT12s_LargeMIMO}. Define power allocation vector as
$\mathbf{p}=\left[p_{1},...,p_{K}\right]^{T}$.%
{} 

Define the normalized channel $\bar{\mathbf{H}}=\mathbf{H}/\sqrt{S}$.
Let $\mathbf{h}_{k}^{\dagger}$ and $\bar{\mathbf{h}}_{k}^{\dagger}$
denote, respectively, the $k^{\textrm{th}}$ row of $\mathbf{H}$
and $\bar{\mathbf{H}}$. Define $\bar{\mathbf{H}}_{k}$ as the matrix
$\bar{\mathbf{H}}$ with the $k^{\textrm{th}}$ row removed, and $\mathbf{P}_{k}\triangleq\textrm{diag}\left(p_{1},...,p_{k-1},p_{k+1},...,p_{K}\right).$
Assume that user $k$ has perfect knowledge of the effective channel
$\mathbf{h}_{k}^{\dagger}\mathbf{f}_{k}$ and the interference-plus-noise
power. The SINR of user $k$ is \cite{Muharar_ICC11_RCIRMT}
\begin{equation}
\gamma_{k}\left(\mathcal{A},\alpha,\mathbf{p}\right)=\frac{p_{k}A_{k}^{2}}{B_{k}+\left(1+A_{k}\right)^{2}},\label{eq:SINR}
\end{equation}
where
\begin{eqnarray*}
A_{k}=\bar{\mathbf{h}}_{k}^{\dagger}\left(\bar{\mathbf{H}}_{k}^{\dagger}\bar{\mathbf{H}}_{k}+\alpha\mathbf{I}_{S}\right)^{-1}\bar{\mathbf{h}}_{k},\:\:\:\:\:\:\:\:\:\:\:\:\:\:\:\:\:\:\:\:\:\:\:\:\:\:\:\:\:\:\:\:\:\:\:\:\:\:\:\:\:\:\:\:\:\:\:\:\:\:\:\:\:\:\\
B_{k}=\bar{\mathbf{h}}_{k}^{\dagger}\left(\bar{\mathbf{H}}_{k}^{\dagger}\bar{\mathbf{H}}_{k}+\alpha\mathbf{I}_{S}\right)^{-1}\bar{\mathbf{H}}_{k}^{\dagger}\mathbf{P}_{k}\bar{\mathbf{H}}_{k}\left(\bar{\mathbf{H}}_{k}^{\dagger}\bar{\mathbf{H}}_{k}+\alpha\mathbf{I}_{S}\right)^{-1}\bar{\mathbf{h}}_{k}.
\end{eqnarray*}
The instantaneous transmit power of the $j^{\textrm{th}}$ selected
antenna is given by
\begin{equation}
P_{\mathcal{A}_{j}}\left(\mathcal{A},\alpha,\mathbf{p}\right)=\frac{1}{S}\mathbf{1}_{j}^{T}\bar{\mathbf{F}}\bar{\mathbf{F}}^{\dagger}\mathbf{1}_{j},\label{eq:Pm}
\end{equation}
where $\bar{\mathbf{F}}=\bar{\mathbf{H}}^{\dagger}\left(\bar{\mathbf{H}}\bar{\mathbf{H}}^{\dagger}+\alpha\mathbf{I}_{K}\right)^{-1}\mathbf{P}^{1/2}$;
and $\mathbf{1}_{j}$ is a $K\times1$ vector whose $j^{\textrm{th}}$
element is $1$ and all other elements are zeros.

\section{Optimization Formulation for Dynamic Antenna Selection\label{sec:Optimization-Formulation-for}}

We consider long-term control policy where the active antenna set
$\mathcal{A}$, the regularization factor $\alpha$, and the power
allocation $\mathbf{p}$ are adaptive to the large scale fading $\mathbf{\Sigma}$
only.
\begin{defn}
[Long-term control policy]A long-term antenna selection, regularization
and power control policy $\Psi=\left\{ \Psi_{\mathcal{A}},\Psi_{\alpha},\Psi_{\mathbf{p}}\right\} $
are mappings from the large scale fading matrix $\mathbf{\Sigma}$
to the active antenna set $\mathcal{A}$, the regularization factor
$\alpha$, and the power allocation $\mathbf{p}$ respectively. Specifically,
$\mathcal{A}$, $\alpha,$ and $\mathbf{p}$ are given by: $\Psi_{\mathcal{A}}\left(\mathbf{\Sigma}\right)$,
$\Psi_{\alpha}\left(\mathbf{\Sigma}\right)$ and $\Psi_{\mathbf{p}}\left(\mathbf{\Sigma}\right)$.
\end{defn}

For technical reasons, we consider a sequence of C-RAN systems indexed
by $S=\left\{ 1,2,...\right\} $. In the $S$-th system, there are
$M=\left\lceil \overline{\beta}S\right\rceil $ distributed transmit
antennas and $K=\left\lceil \beta S\right\rceil $ single-antenna
users, where $\beta\in\left(0,1\right]$ and $\frac{1}{\overline{\beta}}\in\left(0,1\right)$
are constant. Correspondingly, we consider a sequence of long-term
control policies $\left\{ \Psi^{(S)}\right\} $ indexed by $S=\left\{ 1,2,...\right\} $,
and apply the $S$-th control policy $\Psi^{(S)}$ to the $S$-th
system. We restrict our attention to the class of policies that satisfy
the following technical assumptions.
\begin{defn}
[Admissible control policy]A sequence of control policies $\left\{ \Psi^{(S)}\right\} $
is admissible if for each $S$, $\Psi_{\mathcal{A}}^{(S)}$ is a mapping:
$\mathbb{R}_{+}^{K\times M}\mapsto\left\{ \mathcal{A}:\mathcal{A}\subset\left\{ 1,...,M\right\} ,\:\left|\mathcal{A}\right|=S\right\} $,
$\Psi_{\alpha}^{(S)}$ is a mapping: $\mathbb{R}_{+}^{K\times M}\mapsto\left[\alpha_{\textrm{min}},\alpha_{\textrm{max}}\right]$,
and $\Psi_{\mathbf{p}}^{(S)}$ is a mapping: $\mathbb{R}_{+}^{K\times M}\mapsto\left[0,P_{\textrm{max}}\right]^{K}$,
where the constants $P_{\textrm{max}}$, $\alpha_{\textrm{max}}$
and $\alpha_{\textrm{min}}$ $\in\left(0,\infty\right)$.
\end{defn}

The objective of the $S$-th control policy $\Psi^{(S)}$ is to maximize
the conditional average weighted sum-rate for the $S$-th system.
Specifically, given large scale fading matrix $\mathbf{\Sigma}^{(S)}\in\mathbb{R}_{+}^{K\times M}$
and weight vector $\mathbf{w}^{(S)}=\left[w_{k}^{(S)}\right]_{k=1,...,K}\in\mathbb{R}_{+}^{K}$
for the $S$-th system, the long-term control $\Psi^{(S)}\left(\mathbf{\Sigma}^{(S)}\right)$
is given by the solution of the following joint optimization problem
\begin{eqnarray}
\mathcal{P}\left(\mathbf{\Sigma}^{\left(S\right)}\right):\underset{\Psi^{(S)}\left(\mathbf{\Sigma}^{(S)}\right)}{\textrm{max}}\mathcal{I}\left(\Psi^{(S)}\left(\mathbf{\Sigma}^{(S)}\right)|\mathbf{\Sigma}^{(S)}\right)\nonumber \\
\textrm{s.t. }\textrm{E}\left[P_{m}\left(\Psi^{(S)}\left(\mathbf{\Sigma}^{(S)}\right)\right)\left|\mathbf{\Sigma}^{(S)}\right.\right]\leq\frac{\rho_{m}}{S},\:\forall m\in\Psi_{\mathcal{A}}^{(S)}\left(\mathbf{\Sigma}^{(S)}\right),\label{eq:PmCons}
\end{eqnarray}
where for given $\mathbf{\Sigma},\mathbf{w}$ and $\Psi\left(\mathbf{\Sigma}\right)=\left\{ \mathcal{A},\alpha,\mathbf{p}\right\} $,
the conditional average weighted sum-rate is
\begin{equation}
\mathcal{I}\left(\Psi\left(\mathbf{\Sigma}\right)|\mathbf{\Sigma}\right)=\textrm{E}\left[\sum_{k=1}^{K}w_{k}\textrm{log}\left(1+\gamma_{k}\left(\Psi\left(\mathbf{\Sigma}\right)\right)\right)\left|\mathbf{\Sigma}\right.\right],\label{eq:WSR}
\end{equation}
$\gamma_{k}\left(\Psi\left(\mathbf{\Sigma}\right)\right)=\gamma_{k}\left(\mathcal{A},\alpha,\mathbf{p}\right)$
is the SINR in (\ref{eq:SINR}), $P_{m}\left(\Psi\left(\mathbf{\Sigma}\right)\right)=P_{m}\left(\mathcal{A},\alpha,\mathbf{p}\right)$
is the per antenna transmit power in (\ref{eq:Pm}) and $\rho_{m}>0$
is a constant.
\begin{rem}
[Per antenna power constraint]In (\ref{eq:PmCons}), the per antenna
power constraint for the $S$-th system is $\frac{\rho_{m}}{S}$ due
to the following reason. It can be verified that $\textrm{E}\left[P_{m}\left(\Psi^{(S)}\left(\mathbf{\Sigma}^{(S)}\right)\right)\left|\mathbf{\Sigma}^{(S)}\right.\right]=\sum_{k=1}^{K}R_{m,k}^{'}p_{k}=O(1/S)\rightarrow0$
as $S\rightarrow\infty$ under an admissible control policy $\Psi^{(S)}$,
where $R_{m,k}^{'}=O\left(\frac{1}{S^{2}}\right)$ is some coefficient
independent of $\mathbf{p}$. Moreover, the effective channel gain
per user is $O\left(S\right)$. Hence, the per-antenna transmit power
required to support a finite data rate for each user in the $S$-th
system is $O(1/S)$ as $S\rightarrow\infty$ and $O(1)$ power allocation
variables $\left\{ p_{1},...,p_{K}\right\} $ are needed to satisfy
the constraint (\ref{eq:PmCons}). Similar observation is also made
in \cite{Marzetta_SPM12_LargeMIMO,Wagner_TIT12s_LargeMIMO} about
in a large MIMO system with $M$ antennas, a per-antenna power of
$O(1/M)$ is needed to support a finite SINR for each user.
\end{rem}

There are several challenges in solving Problem $\mathcal{P}\left(\mathbf{\Sigma}^{\left(S\right)}\right)$.
First, there is no closed-form expression for the optimization objective
and constraints. Second, the problem is combinatorial w.r.t. the antenna
selection and non-convex w.r.t. the regularization factor and power
allocation. The first challenge is tackled in this section by using
the random matrix theory in \cite{COUILLET_CUP2011_rmtwireless} to
derive asymptotically accurate expressions for the optimization objective
and constraints. The second challenge is tackled in Section \ref{sec:Optimization-Solution}.

To derive DE of the weighted sum-rate and transmit power, we require
the following assumptions.

\begin{assumption}[Boundedness of $\mathbf{\Sigma}^{\left(S\right)},\mathbf{w}^{(S)}$]\label{asm:LSFM}$\:$
\begin{enumerate}
\item \textbf{Uniformly Bounded $\mathbf{\Sigma}^{\left(S\right)}$ w.r.t.
$S$:} Let $\left\{ \mathbf{\Sigma}^{(S)}\in\mathbb{R}_{+}^{K\times M}\right\} $
be a sequence of large scale fading matrices (indexed by $S=\left\{ 1,2,...\right\} $)
such that 
\begin{eqnarray}
\limsup_{S\rightarrow\infty}\sup_{1\leq k\leq K,1\leq m\leq M}\sigma_{km}^{(S)} & < & \infty,\label{eq:LFbound}\\
\liminf_{S\rightarrow\infty}\inf_{1\leq k\leq K,1\leq m\leq M}\sigma_{km}^{(S)} & > & 0,\label{eq:LFlowbound}
\end{eqnarray}
where $\sigma_{km}^{(S)}$ is the element at the $k$-th row and $m$-th
column of $\mathbf{\Sigma}^{(S)}$.
\item \textbf{Uniformly Bounded $Kw_{k}$ w.r.t. $S$:} Let $\left\{ \mathbf{w}^{(S)}\in\mathbb{R}_{+}^{K}\right\} $
be a sequence of weight vectors (indexed by $S=\left\{ 1,2,...\right\} $)
such that
\begin{eqnarray*}
\limsup_{S\rightarrow\infty}\sup_{1\leq k\leq K}Kw_{k}^{(S)} & < & \infty.
\end{eqnarray*}

\end{enumerate}
\end{assumption}

The assumption in (\ref{eq:LFbound}) ensures that the normalized
channel matrix has uniformly bounded spectral norm with probability
1, which is required in the proof of Lemma \ref{lem:AsyWS} and \ref{lem:AsyPm}
later.
\begin{prop}
\label{prop:boundSpnorm}Let $\left\{ \mathbf{\Sigma}^{(S)}\right\} $
be as in Assumption \ref{asm:LSFM}. Let $\bar{\mathbf{H}}^{(S)}$
denote the normalized channel matrix corresponding to $\mathbf{\Sigma}^{(S)}$.
We have \textup{$\limsup_{S\rightarrow\infty}\left\Vert \bar{\mathbf{H}}^{(S)\dagger}\bar{\mathbf{H}}^{(S)}\right\Vert \overset{a.s}{<}\infty$.}
\end{prop}

Please refer to Appendix \ref{sub:Proof-of-PropositionSPbound} for
the proof. 

The assumption in (\ref{eq:LFlowbound}) ensures that $\xi_{l}$ in
(\ref{eq:AkfixEqu}) in Lemma \ref{lem:AsyWS} is bounded away from
zero as $S\rightarrow\infty$. Assumption \ref{asm:LSFM}-2) is to
ensure that $\mathcal{I}\left(\Psi^{(S)}\left(\mathbf{\Sigma}^{(S)}\right)|\mathbf{\Sigma}^{(S)}\right)$
is bounded as $S\rightarrow\infty$.
\begin{lem}
[DE of SINR]\label{lem:AsyWS}Let $\left\{ \mathbf{\Sigma}^{(S)}\right\} $
be as in Assumption \ref{asm:LSFM}. We have $\gamma_{k}\left(\Psi^{(S)}\left(\mathbf{\Sigma}^{(S)}\right)\right)-\bar{\gamma}_{k}\left(\Psi^{(S)}\left(\mathbf{\Sigma}^{(S)}\right)\right)\overset{a.s}{\rightarrow}0$
as $S\rightarrow\infty$, where for given $\mathbf{\Sigma}$ and $\Psi\left(\mathbf{\Sigma}\right)=\left\{ \mathcal{A},\alpha,\mathbf{p}\right\} $,
\begin{equation}
\bar{\gamma}_{k}\left(\Psi\left(\mathbf{\Sigma}\right)\right)=\frac{p_{k}\xi_{k}^{2}}{\frac{1}{S}\sum_{l\neq k}^{K}\left[p_{l}\theta_{kl}/\left(1+\xi_{l}\right)^{2}\right]+\left(1+\xi_{k}\right)^{2}},\label{eq:AsySINR}
\end{equation}
where $\boldsymbol{\xi}=\left[\xi_{1},...,\xi_{K}\right]^{T}\in\mathbb{R}_{+}^{K}$
is the unique solution of
\begin{equation}
\xi_{l}=\frac{1}{S}\sum_{m\in\mathcal{A}}\left[\sigma_{lm}^{2}/f_{m}\left(\boldsymbol{\xi}\right)\right],\: l=1,...,K,\label{eq:AkfixEqu}
\end{equation}
with $f_{m}\left(\boldsymbol{\xi}\right)\triangleq\alpha+\frac{1}{S}\sum_{i=1}^{K}\frac{\sigma_{im}^{2}}{1+\xi_{i}}$,
and $\boldsymbol{\theta}_{k}=\left[\theta_{k1},...,\theta_{kK}\right]^{T}$
is
\begin{eqnarray}
\boldsymbol{\theta}_{k} & \triangleq & \left(\mathbf{I}_{K}-\mathbf{D}\right)^{-1}\mathbf{d}_{k},\: k=1,...,K,\label{eq:ThetaLE}
\end{eqnarray}
with $\mathbf{D}=\left[D_{ln}\right]_{l,n=1,...,K}\in\mathbb{R}^{K\times K}$
and $\mathbf{d}_{k}=\left[d_{kl}\right]_{l=1,...,K}\in\mathbb{R}^{K\times1}$
given by
\begin{eqnarray*}
D_{ln} & = & \frac{1}{S}\sum_{m\in\mathcal{A}}\left[\frac{1}{S}\sigma_{lm}^{2}\sigma_{nm}^{2}/\left(\left(1+\xi_{n}\right)^{2}f_{m}^{2}\left(\boldsymbol{\xi}\right)\right)\right],\\
d_{kl} & = & \frac{1}{S}\sum_{m\in\mathcal{A}}\left[\sigma_{km}^{2}\sigma_{lm}^{2}/f_{m}^{2}\left(\boldsymbol{\xi}\right)\right].
\end{eqnarray*}

\end{lem}

\begin{lem}
[DE of per-antenna transmit power]\label{lem:AsyPm}Let $\left\{ \mathbf{\Sigma}^{(S)}\right\} $
be as in Assumption \ref{asm:LSFM}. We have $SP_{m}\left(\Psi^{(S)}\left(\mathbf{\Sigma}^{(S)}\right)\right)-S\bar{P}_{m}\left(\Psi^{(S)}\left(\mathbf{\Sigma}^{(S)}\right)\right)\overset{a.s}{\rightarrow}0,\forall m\in\Psi_{\mathcal{A}}^{(S)}\left(\mathbf{\Sigma}^{(S)}\right)$
as $S\rightarrow\infty$, where for given $\mathbf{\Sigma}$ and $\Psi\left(\mathbf{\Sigma}\right)=\left\{ \mathcal{A},\alpha,\mathbf{p}\right\} $
\begin{equation}
\bar{P}_{m}\left(\Psi\left(\mathbf{\Sigma}\right)\right)=\frac{\alpha^{-1}}{S^{2}}\psi_{m}^{-2}\left(\mathbf{v}\right)\sum_{i=1}^{K}\sigma_{im}^{2}\left(p_{i}v_{i}-\varphi_{i}\right),\label{eq:AsyPm}
\end{equation}
where $\mathbf{v}=\left[v_{1},...,v_{K}\right]^{T}\in\mathbb{R}_{++}^{K}$
is the unique solution of
\begin{equation}
v_{l}=\frac{1}{\alpha+\frac{1}{S}\sum_{m\in\mathcal{A}}\left[\sigma_{lm}^{2}/\psi_{m}\left(\mathbf{v}\right)\right]},\: l=1,...,K,\label{eq:vfix}
\end{equation}
with $\psi_{m}\left(\mathbf{v}\right)\triangleq1+\frac{1}{S}\sum_{i=1}^{K}\sigma_{im}^{2}v_{i}$,
and $\boldsymbol{\varphi}=\left[\varphi_{1},...,\varphi_{K}\right]^{T}$
is 
\begin{eqnarray}
\boldsymbol{\varphi} & \triangleq & \left(\alpha\mathbf{I}_{K}+\mathbf{\Delta}-\mathbf{C}\right)^{-1}\mathbf{c},\label{eq:FaiDef}
\end{eqnarray}
with $\mathbf{\Delta}=\textrm{diag}\left(\Delta_{1},...,\Delta_{K}\right)$,
$\mathbf{C}=\left[C_{ln}\right]_{l,n=1,...,K}\in\mathbb{R}^{K\times K}$
and $\mathbf{c}=\left[c_{l}\right]_{l=1,...,K}\in\mathbb{R}^{K\times1}$
given by
\begin{eqnarray*}
\Delta_{l} & = & \frac{1}{S}\sum_{m\in\mathcal{A}}\left[\sigma_{lm}^{2}/\psi_{m}\left(\mathbf{v}\right)\right],\: l=1,...,K,\\
C_{ln} & = & \frac{1}{S}\sum_{m\in\mathcal{A}}\left[\frac{1}{S}\sigma_{lm}^{2}\sigma_{nm}^{2}v_{l}/\psi_{m}^{2}\left(\mathbf{v}\right)\right],\\
c_{l} & = & \frac{1}{S}\sum_{m\in\mathcal{A}}\left[\frac{\sigma_{lm}^{2}v_{l}}{\psi_{m}^{2}\left(\mathbf{v}\right)}\left(p_{l}+\frac{1}{S}\sum_{i=1}^{K}\sigma_{im}^{2}v_{i}\left(p_{l}-p_{i}\right)\right)\right].
\end{eqnarray*}

\end{lem}

The proof of Lemma \ref{lem:AsyWS} is similar to that of \cite[Thereom 2]{Wagner_TIT12s_LargeMIMO}
and is omitted for conciseness%
\footnote{In \cite[Thereom 2]{Wagner_TIT12s_LargeMIMO}, $\alpha$ is a positive
constant. However, it can be verified that the proof of \cite[Thereom 2]{Wagner_TIT12s_LargeMIMO}
still holds if we replace the constant $\alpha$ by $\alpha^{(S)}=\Psi_{\alpha}^{(S)}\left(\mathbf{\Sigma}^{(S)}\right)$
in Lemma \ref{lem:AsyWS} due to the bounded constraint on $\Psi_{\alpha}^{(S)}$.%
}. The proof of Lemma \ref{lem:AsyPm} is given in Appendix \ref{sub:Proof-of-mainTheorem}.
\begin{rem}
The equations in (\ref{eq:AkfixEqu}) and (\ref{eq:vfix}) can be
solved using Proposition 1 in \cite{Wagner_TIT12s_LargeMIMO}.\hfill \IEEEQED
\end{rem}

Then the following theorem holds.
\begin{thm}
[Asymptotic equivalence of Problem $\mathcal{P}$]\label{thm:AsyEqP}Let
$\left\{ \mathbf{\Sigma}^{(S)}\right\} ,\left\{ \mathbf{w}^{(S)}\right\} $
be as in Assumption \ref{asm:LSFM} and $\Psi^{(S)*}\left(\mathbf{\Sigma}^{(S)}\right)$
be an optimal solution of the following problem 
\begin{eqnarray}
\mathcal{P}_{E}\left(\mathbf{\Sigma}^{\left(S\right)}\right):\:\underset{\Psi^{(S)}\left(\mathbf{\Sigma}^{(S)}\right)}{\textrm{max}}\:\bar{\mathcal{I}}\left(\Psi^{(S)}\left(\mathbf{\Sigma}^{(S)}\right)|\mathbf{\Sigma}^{(S)}\right)\nonumber \\
\triangleq\sum_{k=1}^{K}w_{k}^{(S)}\textrm{log}\left(1+\bar{\gamma}_{k}\left(\Psi^{(S)}\left(\mathbf{\Sigma}^{(S)}\right)\right)\right)\nonumber \\
\textrm{s.t.}\:\bar{P}_{m}\left(\Psi^{(S)}\left(\mathbf{\Sigma}^{(S)}\right)\right)\leq\frac{\rho_{m}}{S},\:\forall m\in\Psi_{\mathcal{A}}^{(S)}\left(\mathbf{\Sigma}^{(S)}\right).\label{eq:AsyCons}
\end{eqnarray}
Let $\mathcal{I}^{(S)\circ}$ be the optimal value of Problem $\mathcal{P}\left(\mathbf{\Sigma}^{\left(S\right)}\right)$
with weight vector $\mathbf{w}^{(S)}$. Then as $S\rightarrow\infty$,
we have
\begin{eqnarray}
S\textrm{E}\left[P_{m}\left(\Psi^{(S)*}\left(\mathbf{\Sigma}^{(S)}\right)\right)\left|\mathbf{\Sigma}^{(S)}\right.\right]-\rho_{m}\leq0,\nonumber \\
\forall m\in\Psi_{\mathcal{A}}^{(S)*}\left(\mathbf{\Sigma}^{(S)}\right),\label{eq:TS1}\\
\mathcal{I}\left(\Psi^{(S)*}\left(\mathbf{\Sigma}^{(S)}\right)|\mathbf{\Sigma}^{(S)}\right)-\mathcal{I}^{(S)\circ}\rightarrow0.\label{eq:TS3}
\end{eqnarray}

\end{thm}

Please refer to Appendix \ref{sub:Proof-of-Theorem_AsyEqP} for the
proof. Note that in both $\mathcal{P}\left(\mathbf{\Sigma}^{\left(S\right)}\right)$
and $\mathcal{P}_{E}\left(\mathbf{\Sigma}^{\left(S\right)}\right)$,
$\Psi^{(S)}$ must be admissible. 
\begin{rem}
In Appendix \ref{sub:Proof-of-Theorem_AsyEqP}, we prove that for
a given sequence of admissible control policies $\left\{ \Psi^{(S)}\right\} $
and $\left\{ \mathbf{\Sigma}^{(S)}\right\} ,\left\{ \mathbf{w}^{(S)}\right\} $
that satisfy Assumption \ref{asm:LSFM}, the conditional average weighted
sum-rate $\mathcal{I}\left(\Psi^{(S)}\left(\mathbf{\Sigma}^{(S)}\right)|\mathbf{\Sigma}^{(S)}\right)$
(conditioned on the large scale fading matrix $\mathbf{\Sigma}^{(S)}$)
converges to the DE $\bar{\mathcal{I}}\left(\Psi^{(S)}\left(\mathbf{\Sigma}^{(S)}\right)|\mathbf{\Sigma}^{(S)}\right)$
as $S\rightarrow\infty$. In the proposed long-term control policy,
the antenna selection $\mathcal{A}$ is adaptive to the large scale
fading matrix $\mathbf{\Sigma}$ only. Hence, for a given sequence
of $\mathbf{\Sigma}^{(S)}$, the antenna set $\Psi_{\mathcal{A}}^{(S)}\left(\mathbf{\Sigma}^{(S)}\right)$
is deterministic for each $S$ and as a result, the DE convergence
holds true. On the other hand, if $\mathcal{A}$ were adaptive to
the short term CSI $\mathbf{H}$, then conditioned on $\mathbf{\Sigma}^{(S)}$,
$\mathcal{A}$ would still be random and the DE convergence would
fail. Similar conclusion has also been made in \cite{Caire_TIT13_JSDM}
that the DE of the data rate in\textbf{ }massive MIMO system with
user selection is valid as long as the user selection is independent
of the instantaneous CSI $\mathbf{H}$.
\end{rem}

\begin{rem}
For centralized large MIMO downlink, a DE of the SINR has been provided
in Theorem 2 of \cite{Wagner_TIT12s_LargeMIMO} under per-user channel
transmit correlation with correlation matrices $\mathbf{\Theta}_{k}\triangleq\textrm{E}\left[\mathbf{h}_{k}\mathbf{h}_{k}^{\dagger}\right],\:\forall k$
and imperfect CSIT with CSIT errors $\tau_{k}$'s. Our channel model
is a special case of that in \cite{Wagner_TIT12s_LargeMIMO} with%
\footnote{This assumption is reasonable since the correlations between the geographically
distributed antennas are indeed negligible.%
} $\mathbf{\Theta}_{k}=\boldsymbol{\sigma}_{k}^{2}\triangleq\textrm{diag}\left(\sigma_{k\mathcal{A}_{1}}^{2},...,\sigma_{k\mathcal{A}_{S}}^{2}\right),\:\forall k$
and $\tau_{k}=0,\:\forall k$ in \cite{Wagner_TIT12s_LargeMIMO}.
However, this paper and \cite{Wagner_TIT12s_LargeMIMO} focus on different
network topologies (distributed versus centralized). As a result,
there are some new technical challenges:
\begin{itemize}
\item Due to the distributed topology, we have to consider per-antenna power
constraint, which is more complicated than the sum power constraint
considered in \cite{Wagner_TIT12s_LargeMIMO}. For example, in the
DE of the SINR in Theorem 2 of \cite{Wagner_TIT12s_LargeMIMO}, the
RZF precoding matrix is scaled to satisfy the sum power constraint.
However, the per-antenna power constraint cannot be satisfied by simply
scaling the precoding matrix $\mathbf{F}$, and we need to derive
the DE for the per-antenna transmit power as in Lemma \ref{lem:AsyPm}.
Moreover, the per-antenna power constraint has to be explicitly handled
by the optimization algorithm, which makes both the algorithm design
and performance analysis more difficult.
\item The assumptions for the theoretical results are different between
the distributed and the centralized topologies. Theorem 2 of \cite{Wagner_TIT12s_LargeMIMO}
requires the following assumptions: A1) $\mathbf{\bar{H}}^{\dagger}\mathbf{\bar{H}}$
has, almost surely, uniformly bounded spectral norm on $M$. The optimization
problems in \cite{Wagner_TIT12s_LargeMIMO} focus on the case whereby
$\mathbf{\Theta}_{k}=\mathbf{\Theta},\:\forall k$, under which A1
is true. However, it is not clear if A1 holds for the distributed
topology. In this paper, we replace A1 with Assumption \ref{asm:LSFM}-1),
which is a more mild assumption in practical systems. 
\item In Theorem \ref{thm:AsyEqP}, we formally proved the asymptotic equivalence
between Problem $\mathcal{P}$ and its deterministic approximation
$\mathcal{P}_{E}$. The proof of Theorem \ref{thm:AsyEqP} is non-trivial.
However, in \cite{Wagner_TIT12s_LargeMIMO}, the problem formulation
is directly based on the deterministic approximation and there is
no proof of the asymptotic equivalence between the ``original problem''
and its deterministic approximation. 
\item Compared to the optimization problems in \cite{Wagner_TIT12s_LargeMIMO},
problem $\mathcal{P}_{E}$ is much more difficult to solve due to
the heterogeneous path loss and the combinatorial nature of the antenna
selection problem.\hfill \IEEEQED
\end{itemize}
\end{rem}

\section{Optimization Solution for $\mathcal{P}_{E}$\label{sec:Optimization-Solution}}

In Theorem \ref{thm:AsyEqP}, we let $S\rightarrow\infty$ to establish
the asymptotic equivalence between Problem $\mathcal{P}$ and $\mathcal{P}_{E}$.
In this section, we focus on solving $\mathcal{P}_{E}$ for large
but finite $M,K,S$, which is the case for a practical large C-RAN.
By Theorem 1, the optimal value $\mathcal{I}^{*}$ of $\mathcal{P}_{E}\left(\Sigma\right)$
and the optimal value $\mathcal{I}^{\circ}$ of $\mathcal{P}\left(\Sigma\right)$
satisfy $\mathcal{I}^{*}=\mathcal{I}^{\circ}+o(1)$ for fixed $M,K,S$,
where $o(1)\rightarrow0$ as $S\rightarrow\infty$. This implies that
the solution of Problem $\mathcal{P}\left(\Sigma\right)$ can still
be well approximated by the solution of $\mathcal{P}_{E}\left(\Sigma\right)$
for large but finite $M,K,S$. Since we focus on solving $\mathcal{P}_{E}\left(\Sigma\right)$
for given $\Sigma$ in the rest of the paper, we will omit the argument
$\Sigma$ in $\mathcal{P}_{E}$ and explicitly express $\bar{\mathcal{I}}\left(\Psi\left(\mathbf{\Sigma}\right)|\mathbf{\Sigma}\right)$
as $\bar{\mathcal{I}}\left(\Psi\left(\mathbf{\Sigma}\right)\right)=\bar{\mathcal{I}}\left(\mathcal{A},\alpha,\mathbf{p}\right)$,
where $\left\{ \mathcal{A},\alpha,\mathbf{p}\right\} =\Psi\left(\mathbf{\Sigma}\right)$.

\subsection{Problem Decomposition\label{sub:Problem-Decomposition}}

Under an admissible control policy, the power allocation vector is
bounded as $\underset{1\leq k\leq K}{\textrm{max}}\: p_{k}\leq P_{\textrm{max}}$.
We first show that this bounded power constraint can be relaxed in
Problem $\mathcal{P}_{E}$.
\begin{prop}
\label{prop:optpboundr}For fixed $M,K,S$, let $\mathcal{P}_{E}^{'}$
denote a relaxed problem of $\mathcal{P}_{E}$ obtained by removing
the bounded power constraint $\underset{1\leq k\leq K}{\textrm{max}}\: p_{k}\leq P_{\textrm{max}}$
in $\mathcal{P}_{E}$. For sufficiently large $P_{\textrm{max}}$,
the optimal power allocation $\mathbf{p}^{*}=\left[p_{1}^{*},...,p_{K}^{*}\right]$
of $\mathcal{P}_{E}^{'}$ satisfies $\underset{1\leq k\leq K}{\textrm{max}}\: p_{k}^{*}\leq P_{\textrm{max}}$
and thus $\mathcal{P}_{E}$ and $\mathcal{P}_{E}^{'}$ are equivalent.\hfill \IEEEQED
\end{prop}

Please refer to Appendix \ref{sub:Proof-of-Proposition} for the proof.

Using primal decomposition \cite{Boyd_03note_primaldecomp} and Proposition
\ref{prop:optpboundr}, for sufficiently large $P_{\textrm{max}}$,
$\mathcal{P}_{E}$ can be decomposed into the following two subproblems: 

\textbf{Subproblem 1 }(Optimization of $\mathbf{p}$ and $\alpha$
under fixed $\mathcal{A}$):
\begin{equation}
\mathcal{P}_{1}\left(\mathcal{A}\right):\:\underset{\alpha\in\left[\alpha_{\textrm{min}},\alpha_{\textrm{max}}\right],\mathbf{p}\geq\mathbf{0}}{\textrm{max}}\:\bar{\mathcal{I}}\left(\mathcal{A},\alpha,\mathbf{p}\right),\:\textrm{s.t.}\:(\ref{eq:AsyCons})\:\textrm{is}\:\textrm{satisfied}.\label{eq:AsyPfixA}
\end{equation}

\textbf{Subproblem 2 }(Optimization of $\mathcal{A}$): 
\begin{eqnarray}
\mathcal{P}_{2}: & \underset{\mathcal{A}}{\textrm{max}} & \bar{\mathcal{I}}\left(\mathcal{A},\alpha^{*}\left(\mathcal{A}\right),\mathbf{p}^{*}\left(\mathcal{A}\right)\right),\label{eq:SubP3}\\
 & \textrm{s.t.} & \mathcal{A}\subseteq\left\{ 1,...,M\right\} ,\:\textrm{and}\:\left|\mathcal{A}\right|=S,\nonumber 
\end{eqnarray}
where $\alpha^{*}\left(\mathcal{A}\right),\mathbf{p}^{*}\left(\mathcal{A}\right)$
is the optimal solution of $\mathcal{P}_{1}\left(\mathcal{A}\right)$.

Subproblem 1 is non-convex. Although the gradient projection (GP)
method \cite{Bertsekas_book99_NProgramming} is usually used to find
a stationary point for the constrained non-convex problem, it cannot
be applied here because the power constraint functions in (\ref{eq:AsyCons})
are very complicated w.r.t. $\alpha$ and it is very difficult to
calculate the projection of $\alpha$ and $\mathbf{p}$ on the feasible
set of $\mathcal{P}_{1}\left(\mathcal{A}\right)$. In Section \ref{sub:Algorithm-S1-for},
we combine the weighted MMSE (WMMSE) approach in \cite{Luo_TSP11_WMMSE}
and the bisection method to find a stationary point for $\mathcal{P}_{1}\left(\mathcal{A}\right)$.
In Section \ref{sub:Algorithm-S3}, we propose an efficient algorithm
for Subproblem 2. For some special cases discussed in Section \ref{sec:Simplified-Solutions-and},
the proposed algorithms are asymptotically optimal.

\subsection{Solution of Subproblem 1\label{sub:Algorithm-S1-for}}

We first propose an efficient algorithm to solve Subproblem 1 with
fixed $\alpha$, which can be expressed as follows:
\begin{equation}
\mathcal{P}_{1a}\left(\mathcal{A},\alpha\right):\:\underset{\mathbf{p}\geq0}{\textrm{max}}\:\bar{\mathcal{I}}\left(\mathcal{A},\alpha,\mathbf{p}\right),\:\textrm{s.t.}\:(\ref{eq:AsyCons})\:\textrm{is}\:\textrm{satisfied}.\label{eq:AsyPfixruo}
\end{equation}
Then we give the overall solution of Subproblem 1.

\subsubsection{Algorithm S1a for Solving $\mathcal{P}_{1a}\left(\mathcal{A},\alpha\right)$}

$\mathcal{P}_{1a}\left(\mathcal{A},\alpha\right)$ can be rewritten
as a weighted sum-rate maximization problem (WSRMP) under the linear
constraints for $K$-user interference channel as follows. First,
rewrite the objective $\bar{\mathcal{I}}\left(\mathcal{A},\alpha,\mathbf{p}\right)$
as
\[
\bar{\mathcal{I}}\left(\mathcal{A},\alpha,\mathbf{p}\right)=\sum_{k=1}^{K}w_{k}\textrm{log}\left(1+g_{kk}p_{k}/\left(1+\sum_{l\neq k}^{K}g_{kl}p_{l}\right)\right),
\]
\[
g_{kk}\triangleq\frac{\xi_{k}^{2}}{\left(1+\xi_{k}\right)^{2}},\:\forall k,\: g_{kl}\triangleq\frac{\theta_{kl}}{S\left(1+\xi_{l}\right)^{2}\left(1+\xi_{k}\right)^{2}},\:\forall k\neq l.
\]
Define a $K\times S$ matrix $\hat{\mathbf{R}}$ with the elements
given by
\[
\hat{R}_{kj}=\frac{1}{S}\alpha^{-1}\sigma_{k\mathcal{A}_{j}}^{2}\psi_{\mathcal{A}_{j}}^{-2}\left(\mathbf{v}\right),\:\forall k,j,
\]
and define $\bar{\mathbf{R}}$ as a $K\times K$ matrix with each
element given by
\begin{eqnarray*}
\bar{R}_{kk} & = & \frac{1}{S}\sum_{m\in\mathcal{A}}\left[\left(1+\frac{1}{S}\sum_{i\neq k}^{K}\sigma_{im}^{2}v_{i}\right)\sigma_{km}^{2}v_{k}/\psi_{m}^{2}\left(\mathbf{v}\right)\right],\:\forall k,\\
\bar{R}_{kl} & = & -\frac{1}{S}\sum_{m\in\mathcal{A}}\left[\frac{1}{S}\sigma_{lm}^{2}v_{l}\sigma_{km}^{2}v_{k}/\psi_{m}^{2}\left(\mathbf{v}\right)\right],\:\forall k\neq l.
\end{eqnarray*}
Let $\mathbf{V}=\textrm{diag}\left(v_{1},...,v_{K}\right)$. Then
the per antenna power constraint in (\ref{eq:AsyCons}) can be rewritten
as $\mathbf{R}\mathbf{p}\leq\boldsymbol{\rho}$, where $\mathbf{R}\triangleq\hat{\mathbf{R}}^{T}\left[\mathbf{V}-\left(\alpha\mathbf{I}_{K}+\mathbf{\Delta}-\mathbf{C}\right)^{-1}\bar{\mathbf{R}}\right]\in\mathbb{R}^{S\times K}$,
and $\boldsymbol{\rho}=\left[\rho_{\mathcal{A}_{1}},...,\rho_{\mathcal{A}_{S}}\right]^{T}$.

In \cite{Luo_TSP11_WMMSE}, a WMMSE algorithm was proposed to find
a stationary point for the WSRMP in MIMO interfering broadcast channels
under per-BS power constraints. In the following, the WMMSE algorithm
is tailored and generalized to solve $\mathcal{P}_{1a}\left(\mathcal{A},\alpha\right)$
under multiple linear constraints.

Following a similar proof as that of \cite[Theorem 1]{Luo_TSP11_WMMSE},
it can be shown that $\mathbf{p}^{*}\left(\mathcal{A},\alpha\right)=\mathbf{q}^{*}\circ\mathbf{q}^{*}$
is the optimal solution of $\mathcal{P}_{1a}\left(\mathcal{A},\alpha\right)$,
where the notation $\circ$ denotes the Hadamard product; and $\mathbf{q}^{*}$
is the optimal solution of
\begin{equation}
\underset{\mathbf{q},\boldsymbol{\upsilon},\boldsymbol{\omega}}{\textrm{min}}\sum_{k=1}^{K}w_{k}\left(\omega_{k}e_{k}-\textrm{log}\omega_{k}\right),\:\textrm{s.t}.\:\mathbf{R}\left(\mathbf{q}\circ\mathbf{q}\right)\leq\boldsymbol{\rho},\label{eq:WMMSEP}
\end{equation}
where $\mathbf{q},\boldsymbol{\upsilon}$ and $\boldsymbol{\omega}\geq\mathbf{0}$
are vectors in $\mathbb{R}^{K}$; and $e_{k}=\left(1-\upsilon_{k}\sqrt{g_{kk}}q_{k}\right)^{2}+\sum_{l\neq k}\upsilon_{k}^{2}g_{kl}q_{l}^{2}+\upsilon_{k}^{2}$.
Hence we only need to solve Problem (\ref{eq:WMMSEP}), which is convex
in each of the optimization variables $\mathbf{q},\boldsymbol{\upsilon},\boldsymbol{\omega}$.
We can use the block coordinate decent method to solve (\ref{eq:WMMSEP}).
First, for fixed $\mathbf{q},\boldsymbol{\upsilon}$, the optimal
$\boldsymbol{\omega}$ is given by $\omega_{k}^{*}=e_{k}^{-1},\forall k$.
Second, for fixed $\mathbf{q},\boldsymbol{\omega}$, the optimal $\boldsymbol{\upsilon}$
is given by $\upsilon_{k}^{*}=\left(\sum_{l=1}^{K}g_{kl}q_{l}^{2}+1\right)^{-1}\sqrt{g_{kk}}q_{k},\forall k$.
Finally, for fixed $\boldsymbol{\upsilon},\boldsymbol{\omega}$, the
optimal $\mathbf{q}$ is given by the solution of the following optimization
problem:
\begin{eqnarray}
\underset{\mathbf{q}}{\textrm{min}}\sum_{k=1}^{K}\left(w_{k}\omega_{k}\left(1-\upsilon_{k}\sqrt{g_{kk}}q_{k}\right)^{2}+\sum_{l\neq k}w_{l}\omega_{l}\upsilon_{l}^{2}g_{lk}q_{k}^{2}\right)\label{eq:optq}\\
\textrm{s.t.}\:\mathbf{R}\left(\mathbf{q}\circ\mathbf{q}\right)\leq\boldsymbol{\rho}.\:\:\:\:\:\:\:\:\:\:\:\:\:\:\:\:\:\:\:\:\:\:\:\:\:\:\:\:\:\:\:\:\:\:\:\:\:\:\:\:\:\:\:\:\:\:\:\:\:\:\:\:\:\:\nonumber 
\end{eqnarray}
Problem (\ref{eq:optq}) is a convex quadratic optimization problem
which can be solved using the Lagrange dual method. Specifically,
the Lagrange function of Problem (\ref{eq:optq}) is given by
\begin{eqnarray*}
L\left(\boldsymbol{\lambda},\mathbf{q}\right)=\sum_{k=1}^{K}\left(w_{k}\omega_{k}\left(1-\upsilon_{k}\sqrt{g_{kk}}q_{k}\right)^{2}+\sum_{l\neq k}w_{l}\omega_{l}\upsilon_{l}^{2}g_{lk}q_{k}^{2}\right)\\
+\boldsymbol{\lambda}^{T}\left(\mathbf{R}\left(\mathbf{q}\circ\mathbf{q}\right)-\boldsymbol{\rho}\right),\:\:\:\:\:\:\:\:\:\:\:\:\:\:\:\:\:\:\:\:\:\:\:\:\:\:\:\:\:\:\:\:\:\:\:\:\:\:\:\:\:\:\:\:\:\:\:\:\:\:\:\:\:\:
\end{eqnarray*}
where $\boldsymbol{\lambda}\in\mathbb{R}_{+}^{S}$ is the Lagrange
multiplier vector. The dual function of Problem (\ref{eq:optq}) is
\begin{equation}
J\left(\boldsymbol{\lambda}\right)=\underset{\mathbf{q}}{\textrm{min}}L\left(\boldsymbol{\lambda},\mathbf{q}\right).\label{eq:DualFun}
\end{equation}
The minimization problem in (\ref{eq:DualFun}) can be decomposed
into $K$ independent problems as 
\begin{equation}
\underset{q_{k}}{\textrm{min}}\:\left\{ w_{k}\omega_{k}\left(1-\upsilon_{k}\sqrt{g_{kk}}q_{k}\right)^{2}+\left(\sum_{l\neq k}w_{l}\omega_{l}\upsilon_{l}^{2}g_{lk}+\boldsymbol{\lambda}^{T}\mathbf{r}_{k}\right)q_{k}^{2}\right\} ,\label{eq:maxLm}
\end{equation}
where $\mathbf{r}_{k}$ is the $k^{\textrm{th}}$ column of $\mathbf{R}$.
For fixed $\boldsymbol{\lambda}$, Problem (\ref{eq:maxLm}) has a
closed-form solution given by 
\begin{equation}
q_{k}^{*}\left(\boldsymbol{\lambda}\right)=\left(\sum_{l=1}^{K}w_{l}\omega_{l}\upsilon_{l}^{2}g_{lk}+\boldsymbol{\lambda}^{T}\mathbf{r}_{k}\right)^{-1}w_{k}\sqrt{g_{kk}}\upsilon_{k}\omega_{k}.\label{eq:optqk}
\end{equation}
Since (\ref{eq:optq}) is a convex quadratic optimization problem,
the optimal solution is given by $\mathbf{q}_{k}^{*}\left(\tilde{\boldsymbol{\lambda}}\right)=\left[q_{k}^{*}\left(\tilde{\boldsymbol{\lambda}}\right)\right]_{k=1,...,K}$,
where $\tilde{\boldsymbol{\lambda}}$ is the optimal solution of the
dual problem
\begin{equation}
\underset{\boldsymbol{\lambda}}{\textrm{max}}J\left(\boldsymbol{\lambda}\right),\:\textrm{s.t.}\:\boldsymbol{\lambda}\geq\mathbf{0}.\label{eq:mingfun}
\end{equation}
The dual function $J\left(\boldsymbol{\lambda}\right)$ is concave
and it can be verified that $\mathbf{R}\left(\mathbf{q}^{*}\left(\boldsymbol{\lambda}\right)\circ\mathbf{q}^{*}\left(\boldsymbol{\lambda}\right)\right)-\boldsymbol{\rho}$
is a subgradient of $J\left(\boldsymbol{\lambda}\right)$. Hence,
the standard subgradient based methods such as the subgradient algorithm
in \cite{Boyd_03note_Subgradient} or the Ellipsoid method in \cite{Boyd_04Book_Convex_optimization}
can be used to solve Problem (\ref{eq:mingfun}).

The overall algorithm for solving Problem (\ref{eq:WMMSEP}) is summarized
as follows.

\textit{Algorithm S1a} (for solving Problem (\ref{eq:WMMSEP})): 

\textbf{\small{Initialization}}{\small{: Let $\mathbf{q}=c\mathbf{1}$,
where $\mathbf{1}$ denotes a vector of all ones; and $c$ is chosen
such that $\mathbf{R}\left(\mathbf{q}\circ\mathbf{q}\right)\leq\boldsymbol{\rho}$.}}{\small \par}

\textbf{\small{Step 1}}{\small{ Let $\upsilon_{k}=\left(\sum_{l=1}^{K}g_{kl}q_{l}^{2}+1\right)^{-1}\sqrt{g_{kk}}q_{k},\:\forall k$}}{\small \par}

\textbf{\small{Step 2}}{\small{ Let $\omega_{k}=\left(1-\upsilon_{k}\sqrt{g_{kk}}q_{k}\right)^{-1},\:\forall k$}}{\small \par}

\textbf{\small{Step 3}}{\small{ Let $q_{k}=q_{k}^{*}\left(\tilde{\boldsymbol{\lambda}}\right),\:\forall k$;
where $\tilde{\boldsymbol{\lambda}}$ is the optimal solution of (\ref{eq:mingfun})
which can be solved using, e.g., the subgradient algorithm in \cite{Boyd_03note_Subgradient}
or the Ellipsoid method in \cite{Boyd_04Book_Convex_optimization}
with the subgradient of $J\left(\boldsymbol{\lambda}\right)$ given
by $\mathbf{R}\left(\mathbf{q}^{*}\left(\boldsymbol{\lambda}\right)\circ\mathbf{q}^{*}\left(\boldsymbol{\lambda}\right)\right)-\boldsymbol{\rho}$;
and $\mathbf{q}^{*}\left(\boldsymbol{\lambda}\right)$ is given in
(\ref{eq:optqk}).}}{\small \par}

\textbf{\small{Return to Step 1 until convergence.}}\textbf{ }\smallskip{}

The following theorem shows that Algorithm S1a converges to a stationary
point of $\mathcal{P}_{1a}\left(\mathcal{A},\alpha\right)$.
\begin{thm}
[Convergence of Alg. S1a]For any limit point $\left(\tilde{\mathbf{q}},\tilde{\boldsymbol{\upsilon}},\tilde{\boldsymbol{\omega}},\tilde{\boldsymbol{\lambda}}\right)$
of the iterates generated by Algorithm S1a, the corresponding $\tilde{\mathbf{p}}\left(\mathcal{A},\alpha\right)=\tilde{\mathbf{q}}\circ\tilde{\mathbf{q}}$
and $\tilde{\boldsymbol{\lambda}}$ satisfies the KKT conditions of
$\mathcal{P}_{1a}\left(\mathcal{A},\alpha\right)$, which can be expressed
as
\begin{eqnarray}
\nabla_{\mathbf{p}}\bar{\mathcal{I}}\left(\mathcal{A},\alpha,\tilde{\mathbf{p}}\left(\mathcal{A},\alpha\right)\right)-\mathbf{R}^{T}\tilde{\boldsymbol{\lambda}}+\tilde{\boldsymbol{\nu}} & = & 0;\label{eq:KKT}\\
\textrm{diag}\left(\boldsymbol{\rho}\right)\tilde{\boldsymbol{\lambda}}-\textrm{diag}\left(\tilde{\boldsymbol{\lambda}}\right)\mathbf{R}\tilde{\mathbf{p}}\left(\mathcal{A},\alpha\right) & = & 0;\nonumber \\
\textrm{diag}\left(\tilde{\boldsymbol{\nu}}\right)\tilde{\mathbf{p}}\left(\mathcal{A},\alpha\right) & = & 0;\nonumber 
\end{eqnarray}
where $\tilde{\boldsymbol{\lambda}}$ and $\tilde{\boldsymbol{\nu}}=\mathbf{R}^{T}\tilde{\boldsymbol{\lambda}}-\nabla_{\mathbf{p}}\bar{\mathcal{I}}\left(\mathcal{A},\alpha,\tilde{\mathbf{p}}\left(\mathcal{A},\alpha\right)\right)$
are the Lagrange multipliers associated with the constraints $\mathbf{R}\mathbf{p}\leq\boldsymbol{\rho}$
and $\mathbf{p}\geq0$ respectively; and $\nabla_{\mathbf{p}}\bar{\mathcal{I}}\left(\mathcal{A},\alpha,\tilde{\mathbf{p}}\left(\mathcal{A},\alpha\right)\right)=\left[\frac{\partial\bar{\mathcal{I}}}{\partial p_{1}},...,\frac{\partial\bar{\mathcal{I}}}{\partial p_{K}}\right]$
with 
\[
\frac{\partial\bar{\mathcal{I}}}{\partial p_{k}}=\frac{w_{k}g_{kk}}{\left(\tilde{\Omega}_{k}+g_{kk}\tilde{p}_{k}\left(\mathcal{A},\alpha\right)\right)}-\sum_{l\neq k}^{K}\frac{w_{l}g_{lk}g_{ll}\tilde{p}_{l}\left(\mathcal{A},\alpha\right)}{\tilde{\Omega}_{l}\left(\tilde{\Omega}_{l}+g_{ll}\tilde{p}_{l}\left(\mathcal{A},\alpha\right)\right)}.
\]
where
\begin{equation}
\tilde{\Omega}_{k}=1+\sum_{l\neq k}^{K}g_{kl}\tilde{p}_{l}\left(\mathcal{A},\alpha\right),\:\forall k.\label{eq:omgl}
\end{equation}
 
\end{thm}

\begin{IEEEproof}
Following a similar proof as that of \cite[Theorem 3]{Luo_TSP11_WMMSE},
we can show that Algorithm S1a converges to a stationary point $\tilde{\mathbf{q}},\tilde{\boldsymbol{\upsilon}},\tilde{\boldsymbol{\omega}}$
of Problem (\ref{eq:WMMSEP}). At the stationary point, the corresponding
$\tilde{\mathbf{q}},\tilde{\boldsymbol{\upsilon}},\tilde{\boldsymbol{\omega}},\tilde{\boldsymbol{\lambda}}$
must satisfy $\textrm{diag}\left(\boldsymbol{\rho}\right)\tilde{\boldsymbol{\lambda}}-\textrm{diag}\left(\tilde{\boldsymbol{\lambda}}\right)\mathbf{R}\left(\tilde{\mathbf{q}}\circ\tilde{\mathbf{q}}\right)=0$
and ({\small{\ref{eq:optqk}}}) with $\tilde{\upsilon}_{k}=\left(\sum_{l=1}^{K}g_{kl}\tilde{q}_{l}^{2}+1\right)^{-1}\sqrt{g_{kk}}\tilde{q}_{k}$,
and $\tilde{\omega}_{k}=\left(1-\tilde{\upsilon}_{k}\sqrt{g_{kk}}\tilde{q}_{k}\right)^{-1}$.
Using the above fact, it can be verified by a direct calculation that
$\tilde{\mathbf{p}}\left(\mathcal{A},\alpha\right)$ and $\tilde{\boldsymbol{\lambda}}$
satisfies the KKT conditions in (\ref{eq:KKT}).
\end{IEEEproof}

\subsubsection{Overall Solution for Subproblem 1\label{sub:Bisection-subP2}}

The following theorem characterizes the overall solution of subproblem
1. 
\begin{thm}
[Stationary point of $\mathcal{P}_{1}\left(\mathcal{A}\right)$]\label{thm:Equivalent-problem-ofP2}Let
$\tilde{\mathbf{p}}\left(\mathcal{A},\alpha\right)$ denote the stationary
point of \textup{$\mathcal{P}_{1a}\left(\mathcal{A},\alpha\right)$}
found by Algorithm S1a. Assume that $\tilde{\mathbf{p}}\left(\mathcal{A},\alpha\right)$
is differentiable over $\alpha$ and define a function 
\begin{equation}
\hat{\mathcal{I}}\left(\mathcal{A},\alpha\right)\triangleq\bar{\mathcal{I}}\left(\mathcal{A},\alpha,\tilde{\mathbf{p}}\left(\mathcal{A},\alpha\right)\right).\label{eq:IheadAruo}
\end{equation}
Then the following are true:
\begin{enumerate}
\item If $\frac{\partial\hat{\mathcal{I}}\left(\mathcal{A},\alpha_{\textrm{min}}\right)}{\partial\alpha}<0$,
then $\alpha_{\textrm{min}},\tilde{\mathbf{p}}\left(\mathcal{A},\alpha_{\textrm{min}}\right)$
is a stationary point of $\mathcal{P}_{1}\left(\mathcal{A}\right)$.
\item If $\frac{\partial\hat{\mathcal{I}}\left(\mathcal{A},\alpha_{\textrm{max}}\right)}{\partial\alpha}>0$,
then $\alpha_{\textrm{max}},\tilde{\mathbf{p}}\left(\mathcal{A},\alpha_{\textrm{max}}\right)$
is a stationary point of $\mathcal{P}_{1}\left(\mathcal{A}\right)$.
\item If $\frac{\partial\hat{\mathcal{I}}\left(\mathcal{A},\alpha_{\textrm{min}}\right)}{\partial\alpha}>0$
and $\frac{\partial\hat{\mathcal{I}}\left(\mathcal{A},\alpha_{\textrm{max}}\right)}{\partial\alpha}<0$,
let $\tilde{\alpha}\left(\mathcal{A}\right)$ be a solution of the
following equation
\begin{equation}
\frac{\partial\hat{\mathcal{I}}\left(\mathcal{A},\alpha\right)}{\partial\alpha}=0,\:\alpha\in\left[\alpha_{\textrm{min}},\alpha_{\textrm{max}}\right],\label{eq:gamdruo}
\end{equation}
i.e., $\tilde{\alpha}\left(\mathcal{A}\right)$ is a stationary point
of $\hat{\mathcal{I}}\left(\mathcal{A},\alpha\right)$. Then $\tilde{\alpha}\left(\mathcal{A}\right),\tilde{\mathbf{p}}\left(\mathcal{A},\tilde{\alpha}\left(\mathcal{A}\right)\right)$
is a stationary point of $\mathcal{P}_{1}\left(\mathcal{A}\right)$. 
\end{enumerate}
\end{thm}

\begin{IEEEproof}
It is easy to see that if $\frac{\partial\hat{\mathcal{I}}\left(\mathcal{A},\alpha_{\textrm{min}}\right)}{\partial\alpha}<0$
($\frac{\partial\hat{\mathcal{I}}\left(\mathcal{A},\alpha_{\textrm{max}}\right)}{\partial\alpha}>0$),
$\alpha_{\textrm{min}}$ ($\alpha_{\textrm{max}}$) is a local maximum
(and thus stationary point) of the problem $\max_{\alpha\in\left[\alpha_{\textrm{min}},\alpha_{\textrm{max}}\right]}\hat{\mathcal{I}}\left(\mathcal{A},\alpha\right)$.
Then Theorem \ref{thm:Equivalent-problem-ofP2} can be proved using
the facts that $\tilde{\mathbf{p}}\left(\mathcal{A},\alpha\right)$
($\alpha=\alpha_{\textrm{min}},\alpha_{\textrm{max}}$ or $\tilde{\alpha}\left(\mathcal{A}\right)$
depending on different cases) is a stationary point of $\mathcal{P}_{1a}\left(\mathcal{A},\alpha\right)$
and $\alpha$ is a stationary point of $\max_{\alpha\in\left[\alpha_{\textrm{min}},\alpha_{\textrm{max}}\right]}\hat{\mathcal{I}}\left(\mathcal{A},\alpha\right)$.
The details are omitted for conciseness.
\end{IEEEproof}

Motivated by Theorem \ref{thm:Equivalent-problem-ofP2}, we propose
the following bisection algorithm to solve $\mathcal{P}_{1}\left(\mathcal{A}\right)$
.

\smallskip{}

\textit{Algorithm S1b} (Bisection search for solving $\mathcal{P}_{1}\left(\mathcal{A}\right)$): 

\textbf{\small{Initialization}}{\small{: If $\frac{\partial\hat{\mathcal{I}}\left(\mathcal{A},\alpha_{\textrm{min}}\right)}{\partial\alpha}<0$,
terminate the algorithm and output $\alpha_{\textrm{min}},\tilde{\mathbf{p}}\left(\mathcal{A},\alpha_{\textrm{min}}\right)$.
If $\frac{\partial\hat{\mathcal{I}}\left(\mathcal{A},\alpha_{\textrm{max}}\right)}{\partial\alpha}>0$,
terminate the algorithm and output $\alpha_{\textrm{max}},\tilde{\mathbf{p}}\left(\mathcal{A},\alpha_{\textrm{max}}\right)$.
Otherwise, choose proper $\alpha_{a},\alpha_{b}$ such that $0<\alpha_{a}<\alpha_{b}$
and $\frac{\partial\hat{\mathcal{I}}\left(\mathcal{A},\alpha_{a}\right)}{\partial\alpha}>0,\:\frac{\partial\hat{\mathcal{I}}\left(\mathcal{A},\alpha_{b}\right)}{\partial\alpha}<0$.}}{\small \par}

\textbf{\small{Step 1}}{\small{: Let $\alpha=\left(\alpha_{a}+\alpha_{b}\right)/2$.
If $\frac{\partial\hat{\mathcal{I}}\left(\mathcal{A},\alpha\right)}{\partial\alpha}\leq0$,
let $\alpha_{b}=\alpha$. Otherwise, let $\alpha_{a}=\alpha$.}}{\small \par}

\textbf{\small{Step 2}}{\small{: If }}\textbf{\small{$\alpha_{b}-\alpha_{a}$
}}{\small{is small enough, terminate the algorithm and output $\alpha,\tilde{\mathbf{p}}\left(\mathcal{A},\alpha\right)$.
Otherwise, return to Step 1.}}{\small \par}
\begin{rem}
It is observed in the simulations that one can always choose sufficiently
large $\alpha_{\textrm{max}}$ and sufficiently small $\alpha_{\textrm{min}}>0$
such that $\frac{\partial\hat{\mathcal{I}}\left(\mathcal{A},\alpha_{\textrm{min}}\right)}{\partial\alpha}>0$
and $\frac{\partial\hat{\mathcal{I}}\left(\mathcal{A},\alpha_{\textrm{max}}\right)}{\partial\alpha}<0$.
Then the constraint $\alpha_{\textrm{min}}\leq\alpha\leq\alpha_{\textrm{max}}$
is never active at the solution found by Algorithm S1b.
\end{rem}

The calculation of $\frac{\partial\hat{\mathcal{I}}\left(\mathcal{A},\alpha\right)}{\partial\alpha}$
in Algorithm S1b is non-trivial due to the lack of analytical expression
for $\hat{\mathcal{I}}\left(\mathcal{A},\alpha\right)$. In the following,
we show how to calculate $\frac{\partial\hat{\mathcal{I}}\left(\mathcal{A},\alpha\right)}{\partial\alpha}$
from the output of Algorithm S1a: $\tilde{\mathbf{p}}\left(\mathcal{A},\alpha\right)$
and $\tilde{\boldsymbol{\lambda}}$. Assuming that $\frac{\partial\tilde{\mathbf{p}}\left(\mathcal{A},\alpha\right)}{\partial\alpha}$,
$\frac{\partial\tilde{\boldsymbol{\lambda}}}{\partial\alpha}$ and
$\frac{\partial\tilde{\boldsymbol{\nu}}}{\partial\alpha}$ exist and
taking partial derivative of the equations in (\ref{eq:KKT}) with
respect to $\alpha$, we obtain a linear equation with $\frac{\partial\tilde{\mathbf{p}}\left(\mathcal{A},\alpha\right)}{\partial\alpha}$,
$\frac{\partial\tilde{\boldsymbol{\lambda}}}{\partial\alpha}$ and
$\frac{\partial\tilde{\boldsymbol{\nu}}}{\partial\alpha}$ as the
variables. Then we can calculate $\frac{\partial\tilde{\mathbf{p}}\left(\mathcal{A},\alpha\right)}{\partial\alpha}$
by solving this linear equation. Finally, the derivative $\frac{\partial\hat{\mathcal{I}}\left(\mathcal{A},\alpha\right)}{\partial\alpha}$
can be calculated as
\begin{eqnarray}
\frac{\partial\hat{\mathcal{I}}\left(\mathcal{A},\alpha\right)}{\partial\alpha} & = & \sum_{k=1}^{K}w_{k}\left(\frac{\sum_{l=1}^{K}\left(\tilde{p}_{l}\left(\mathcal{A},\alpha\right)\frac{\partial g_{kl}}{\partial\alpha}+g_{kl}\frac{\partial\tilde{p}_{l}\left(\mathcal{A},\alpha\right)}{\partial\alpha}\right)}{g_{kk}\tilde{p}_{k}\left(\mathcal{A},\alpha\right)+\tilde{\Omega}_{k}}\right.\nonumber \\
 &  & \left.-\frac{\sum_{l\neq k}^{K}\left(\tilde{p}_{l}\left(\mathcal{A},\alpha\right)\frac{\partial g_{kl}}{\partial\alpha}+g_{kl}\frac{\partial\tilde{p}_{l}\left(\mathcal{A},\alpha\right)}{\partial\alpha}\right)}{\tilde{\Omega}_{k}}\right).\label{eq:Idruo}
\end{eqnarray}
The detailed calculations for $\frac{\partial\tilde{\mathbf{p}}\left(\mathcal{A},\alpha\right)}{\partial\alpha}$,
$\frac{\partial g_{kl}}{\partial\alpha}$'s and $\frac{\partial\hat{\mathcal{I}}\left(\mathcal{A},\alpha\right)}{\partial\alpha}$
can be found in Appendix \ref{sub:Calculation-of-DIdruo}.

\subsection{Algorithm S2 for Solving Subproblem 2\label{sub:Algorithm-S3}}

Subproblem 2 is a combinatorial problem and the optimal solution requires
exhaustive search. We shall propose a low complexity algorithm which
is asymptotically optimal for large $M$ as will be shown in Corollary
\ref{cor:Asymptotic-Optimality-ofS3}. 

Based on the insight obtained in Example 1 and 2, we propose an efficient
algorithm S2 for $\mathcal{P}_{2}$. In step 1, the algorithm selects
antennas that have a direct link with a single user and do not cause
strong interference to others%
\footnote{The phrase ``an antenna causes interference to a user'' refers to
the case when an antenna causes strong interference to a user before
precoding and joint transmission using RZF does not provide much gain
due to some other weak cross links as shown in Example \ref{exa:Strong-cross-link-causes}.%
}. In step 2, the algorithm selects antennas that have strong links
with several users. These antennas have the potential to provide large
cooperative gain. Note that ``bad'' antennas that cause strong interference
may also be selected; however, they will be deleted in step 4. In
step 3, the algorithm selects antennas that have a strong cross link
with a single user and do not cause strong interference to others.
In step 4, a greedy search is performed to replace the ``bad'' antennas
with ``good'' ones from a candidate antenna set $\Gamma_{j}$, which
is carefully chosen to reduce the number of weighted sum-rate calculations
in step 4 as well as to maintain a good performance.

We first define some notations and then give the detailed steps of
Algorithm S2. Let $\tilde{m}_{k}=\textrm{argmax}_{m}\sigma_{km}^{2},$
$k=1,...,K$. Define $\bar{g}_{k}^{d}=\sigma_{k\tilde{m_{k}}}^{2}$.
For $k=1,...,K$, and $m=1,...,M$, let $G_{km}=1$, if $\sigma_{km}^{2}\geq\kappa\bar{g}_{k}^{d},$
and otherwise, let $G_{km}=0$, where $\kappa\in\left(0,1\right)$.
Roughly speaking, $G_{km}$ is an indication of whether antenna $m$
contributes significantly to the communication of user $k$. Simulations
show that the performance of Algorithm S2 is not sensitive to the
choice of $\kappa$ for $\kappa\in\left[\frac{1}{4},\frac{1}{2}\right]$.
Define $\mathcal{K}_{m}=\left\{ k:\: G_{km}=1\right\} ,\: m=1,...,M$.
Let $\tilde{\mathcal{I}}_{\mathcal{A}}\triangleq\bar{\mathcal{I}}\left(\mathcal{A},\tilde{\rho}\left(\mathcal{A}\right),\tilde{\mathbf{p}}\left(\mathcal{A},\tilde{\rho}\left(\mathcal{A}\right)\right)\right)$
denote the optimized weighted sum-rate under $\mathcal{A}$. For any
set of antennas $\mathcal{A}\subseteq\left\{ 1,...,M\right\} $, let
$\bar{\mathcal{A}}$ denote the relative complement of $\mathcal{A}$. 

\smallskip{}

\textit{Algorithm S2} (for solving Subproblem 2):

\textbf{\small{Initialization}}{\small{: Let $\mathcal{A}=\Phi$,
where $\Phi$ denotes the void set.}}{\small \par}

\textbf{\small{Step 1}}{\small{ (Select antennas with a direct link
and no cross link): }}{\small \par}

{\small{$\;$$\;$}}\textbf{\small{For}}{\small{ $k=1$ }}\textbf{\small{to}}{\small{
$K$, if $\left|\mathcal{K}_{\tilde{m}_{k}}\right|=1$, let $\mathcal{A}=\mathcal{A}\cup\tilde{m}_{k}$.
If $\left|\mathcal{A}\right|=S$, go to step 4.}}{\small \par}

\textbf{\small{Step 2}}{\small{ (Select antennas with multiple strong
links): }}{\small \par}

{\small{$\;$$\;$Let $\bar{G}_{m}=\left|\mathcal{K}_{m}\right|B+\sum_{k=1}^{K}\sigma_{km}^{2}$,
where $B$ can be any constant larger than $\underset{1\leq m\leq M}{\textrm{max}}\sum_{k=1}^{K}\sigma_{km}^{2}$. }}{\small \par}

{\small{$\;$$\;$Let $m^{*}=\underset{m\in\bar{\mathcal{A}}}{\textrm{argmax}\:}\bar{G}_{m}$}}{\small \par}

{\small{$\;$$\;$}}\textbf{\small{While}}{\small{ $\left|\mathcal{K}_{m^{*}}\right|\geq2$
and $\left|\mathcal{A}\right|<S$}}{\small \par}

{\small{$\;$$\;$$\;$$\;$Let $\mathcal{A}=\mathcal{A}\cup m^{*}$
and $m^{*}=\underset{m\in\bar{\mathcal{A}}}{\textrm{argmax}\:}\bar{G}_{m}$.}}{\small \par}

{\small{$\;$$\;$}}\textbf{\small{End}}{\small \par}

{\small{$\;$$\;$If $\left|\mathcal{A}\right|=S$, go to step 4.}}{\small \par}

\textbf{\small{Step 3}}{\small{ (Select antennas with a single strong
link): }}{\small \par}

{\small{$\;$$\;$Let $\tilde{k}_{m}=\underset{k}{\textrm{argmax}}\:\sigma_{km}^{2}$
and $I_{m}=w_{\tilde{k}_{m}}\textrm{log}\left(1+\sigma_{\tilde{k}_{m}m}^{2}\right)$.}}{\small \par}

{\small{$\;$$\;$}}\textbf{\small{While}}{\small{ $\left|\mathcal{A}\right|<S$}}{\small \par}

{\small{$\;$$\;$$\;$$\;$Let $m^{*}=\underset{m\in\bar{\mathcal{A}}}{\textrm{argmax}}\: I_{m}$
and $\mathcal{A}=\mathcal{A}\cup m^{*}$.}}{\small \par}

{\small{$\;$$\;$}}\textbf{\small{End}}{\small \par}

\textbf{\small{Step 4}}{\small{ (Greedy search for replacing \textquotedbl{}bad\textquotedbl{}
antennas with \textquotedbl{}good\textquotedbl{} ones): }}{\small \par}

{\small{$\;$$\;$}}\textbf{\small{For}}{\small{ $j=1$ }}\textbf{\small{to}}{\small{
$S$}}{\small \par}

{\small{$\;$$\;$$\;$$\;$Let $\mathcal{A}_{-j}=\mathcal{A}/\mathcal{A}_{j}$.
Let $n^{*}=\underset{m\in\bar{\mathcal{A}}\cap\left\{ m:\left|\mathcal{K}_{m}\right|=1\right\} }{\textrm{argmax}}I_{m}$
and $\Gamma_{j}^{a}=\bar{\mathcal{A}}\cap\left\{ m:\:\left|\mathcal{K}_{m}\right|\geq2\right\} $.}}{\small \par}

{\small{$\;$$\;$$\;$$\;$If $I_{n^{*}}\geq I_{\mathcal{A}_{j}}$
or $\left|\mathcal{K}_{\mathcal{A}_{j}}\right|>1$, let $\Gamma_{j}=\Gamma_{j}^{a}\cup n^{*}$;
otherwise, let $\Gamma_{j}=\Gamma_{j}^{a}$. }}{\small \par}

{\small{$\;$$\;$$\;$$\;$Let $m^{*}=\underset{m\in\Gamma_{j}}{\textrm{argmax}}\:\tilde{\mathcal{I}}_{\mathcal{A}_{-j}\cup m}$. }}{\small \par}

{\small{$\;$$\;$$\;$$\;$If $\tilde{\mathcal{I}}_{\mathcal{A}_{-j}\cup m^{*}}>\tilde{\mathcal{I}}_{\mathcal{A}}$,
let $\mathcal{A}_{j}=m^{*}$.}}{\small \par}

{\small{$\;$$\;$}}\textbf{\small{End}}\smallskip{}

Finally, we elaborate the choice of the candidate antenna set $\Gamma_{j}$
in step 4. The $I_{m}$ defined in step 3 reflects the contribution
of antenna $m$ to the weighted rate of a single user. Then $n^{*}$
in step 4 is the unselected antenna which is likely to contribute
the most to the weighted rate of a single user without causing strong
interference to other users. If $I_{n^{*}}\geq I_{\mathcal{A}_{j}}$,
$n^{*}$ is added in $\Gamma_{j}$. Even if $I_{n^{*}}<I_{\mathcal{A}_{j}}$,
we still add $n^{*}$ in $\Gamma_{j}$ if $\mathcal{A}_{j}$ has the
potential to cause large interference, i.e., $\left|\mathcal{K}_{\mathcal{A}_{j}}\right|>1$.
$\Gamma_{j}^{a}$ are the set of unselected antennas which have the
potential to provide cooperative gain, and are also added in $\Gamma_{j}$.

\section{Structural Solution for Some Special Cases\label{sec:Simplified-Solutions-and}}

\subsection{Large MIMO Network with Collocated Antennas }

We first study the case where the antennas are collocated at the base
station. Specifically, this corresponds to the case where all antennas
experience the same large scale fading: $\sigma_{km}^{2}=\sigma_{k1}^{2},\: k=1,...,K,\: m=1,...,M$.
In this case, any subset $\mathcal{A}$ of $S$ antennas is optimal
for the antenna selection problem since all antennas are statistically
equivalent. We focus on deriving the structural properties of $\mathbf{p}^{*}$
and $\alpha^{*}$. 

We first obtain simpler expressions for asymptotic SINR and transmit
power for $\mathcal{P}_{1}\left(\mathcal{A}\right)$.
\begin{thm}
[DE for collocated antennas]\label{thm:AsyWSCA}Let $\mathbf{\Sigma},\alpha$
be as in Assumption \ref{asm:LSFM}. The following statements are
true for the special case of collocated antennas ($\sigma_{km}^{2}=\sigma_{k1}^{2},\: k=1,...,K,\: m=1,...,M$):
\begin{enumerate}
\item The system of equation:
\begin{equation}
u=\frac{1}{\alpha+\frac{1}{S}\sum_{i=1}^{K}\frac{\sigma_{i1}^{2}}{1+\sigma_{i1}^{2}u}},\label{eq:Defu}
\end{equation}
admits a unique solution $u$ in $\mathbb{R}_{++}$. 
\item Define 
\[
F_{12}=\frac{1}{S}\sum_{i=1}^{K}\frac{\sigma_{i1}^{2}}{\left(1+\sigma_{i1}^{2}u\right)^{2}},\:\bar{F}_{12}\left(\mathbf{p}\right)=\frac{1}{S}\sum_{i=1}^{K}\frac{p_{i}\sigma_{i1}^{2}}{\left(1+\sigma_{i1}^{2}u\right)^{2}}.
\]
As $S\overset{K=S\beta}{\longrightarrow}\infty$, $\gamma_{k}\left(\mathcal{A},\alpha,\mathbf{p}\right)$
in (\ref{eq:SINR}) and $P_{m}\left(\mathcal{A},\alpha,\mathbf{\mathbf{p}}\right)$
in (\ref{eq:Pm}) converge, respectively, to the following deterministic
values almost surely
\begin{eqnarray}
\bar{\gamma}_{k}\left(\mathcal{A},\alpha,\mathbf{P}\right) & = & \frac{p_{k}\sigma_{k1}^{4}u^{2}\left(\alpha+F_{12}\right)}{\bar{F}_{12}\left(\mathbf{p}\right)\sigma_{k1}^{2}u+\left(1+\sigma_{k1}^{2}u\right)^{2}\left(\alpha+F_{12}\right)},\nonumber \\
\bar{P}_{m}\left(\mathcal{A},\alpha,\mathbf{\mathbf{p}}\right) & = & \frac{\bar{F}_{12}\left(\mathbf{p}\right)u}{S\left(\alpha+F_{12}\right)},\:\forall m\in\mathcal{A}.\label{eq:PmCA}
\end{eqnarray}

\end{enumerate}
\end{thm}

The proof is similar to the proof in Appendix \ref{sub:Proof-of-mainTheorem}.%
{} 

Using Theorem \ref{thm:AsyWSCA}, $\mathcal{P}_{1a}\left(\mathcal{A},\alpha\right)$
can be reformulated into a simpler form as follows. First, according
to (\ref{eq:PmCA}), all antennas always have the same transmit power.
Furthermore, it can be verified that the per antenna power constraint
must be achieved with equality at the optimal solution. Combining
these facts and the asymptotic expressions in Theorem \ref{thm:AsyWSCA},
$\mathcal{P}_{1a}\left(\mathcal{A},\alpha\right)$ is equivalent to
the following optimization problem
\begin{equation}
\underset{\mathbf{p}\geq0}{\textrm{max}}\sum_{k=1}^{K}w_{k}\textrm{log}\left(1+\frac{p_{k}\sigma_{k1}^{4}u^{2}}{\sigma_{k1}^{2}P_{T}+\left(1+\sigma_{k1}^{2}u\right)^{2}}\right),\:\textrm{s.t.}\:\frac{\bar{F}_{12}\left(\mathbf{p}\right)u}{\left(\alpha+F_{12}\right)}\leq P_{T},\label{eq:AsyMainPCA}
\end{equation}
where $P_{T}=\underset{m\in\mathcal{A}}{\textrm{min}}\rho_{m}$.

\subsubsection{Water-filling Structure of the Optimal Power Allocation}

For fixed $\mathcal{A},\alpha$, the optimal power allocation $\mathbf{p}^{*}\left(\mathcal{A},\alpha\right)=\left[p_{1}^{*},...,p_{K}^{*}\right]$
is given by:
\begin{equation}
p_{k}^{*}=\left(\frac{w_{k}S\left(1+\sigma_{k1}^{2}u\right)^{2}\left(\alpha+F_{12}\right)}{\lambda\sigma_{k1}^{2}u}-\frac{\sigma_{k1}^{2}P_{T}+\left(1+\sigma_{k1}^{2}u\right)^{2}}{\sigma_{k1}^{4}u^{2}}\right)^{+},\label{eq:pkCA}
\end{equation}
where $\lambda$ is chosen such that $\bar{F}_{12}\left(\mathbf{p}^{*}\left(\mathcal{A},\alpha\right)\right)u/\left(\alpha+F_{12}\right)=P_{T}$.

\subsubsection{Properties of the optimal $\alpha$ in High SNR Regime\label{sub:Asymptotic-Analysis-CA}}

The following theorem summarizes the structural properties of the
optimal solution $\alpha^{*}$ for $\mathcal{P}_{1}\left(\mathcal{A}\right)$.
\begin{thm}
[Properties of $\alpha^{*}$ at high SNR]\label{thm:OptruoCA}For
fixed $K,S$ and sufficiently small $\alpha_{\textrm{min}}$, the
following are true:
\begin{enumerate}
\item $\alpha^{*}=O\left(\frac{1}{P_{T}}\right)$ for large $P_{T}$. 
\item There exists a small enough constant $\alpha_{t}>\alpha_{\textrm{min}}$
such that $\bar{\mathcal{I}}\left(\mathcal{A},\alpha,\mathbf{p}^{*}\left(\mathcal{A},\alpha\right)\right)$,
is a concave function of $\alpha$ for all $\alpha\in\left[\alpha_{\textrm{min}},\alpha_{t}\right)$.
\end{enumerate}
\end{thm}

The proof is given in Appendix \ref{sub:Proof-of-TheoremOptruoCA}.
Theorem \ref{thm:OptruoCA} implies that with sufficiently small $\alpha_{\textrm{min}}$
and initial $\alpha_{b}>\alpha_{a}\geq\alpha_{\textrm{min}}$, Algorithm
S1b will converge to the optimal $\alpha^{*}$ at high enough SNR.
\begin{rem}
[Large MIMO with Collocated Users]When all users are collocated in
the large MIMO network (i.e., the users are very close geographically),
they experience the same large scale fading: $\sigma_{km}^{2}=\sigma_{1m}^{2},\: k=1,...,K,\: m=1,...,M$.
As a result, the optimal active antenna set $\mathcal{A}^{*}$ contains
the antennas that have the largest $S$ large scale fading factors
with the users. Similarly, it can be shown that the optimal power
allocation for $\mathcal{P}_{1a}\left(\mathcal{A},\alpha\right)$
is also a \textit{water-filling solution}. The details are omitted
due to limited space.
\end{rem}

\subsection{Very Large Distributed MIMO Network}

In this section, we derive the asymptotic performance for very large
distributed MIMO networks. Note that the analysis in this section
does not rely on Assumption \ref{asm:LSFM}. The results in this section
are derived only under the large distributed MIMO setup defined below.
\begin{defn}
[Very Large Distributed MIMO Network]\label{def:VLDAS}The coverage
area is a square with side length $R_{c}$. There are $M=N^{2}$ antennas
evenly distributed in the square grid for some integer $N$. The locations
of the $K$ users are randomly generated from a uniform distribution
within the square. The path loss model is given by $\sigma_{km}^{2}=G_{0}r_{km}^{-\zeta},$
where $G_{0}>0$ is a constant, $r_{km}$ is the distance between
the $m^{\textrm{th}}$ antenna and $k^{\textrm{th}}$ user, $\zeta$
is the path loss factor.\hfill \IEEEQED
\end{defn}

Let $\tilde{m}_{k}=\underset{m}{\textrm{argmax}\:}\sigma_{km}^{2}$.
Define $\bar{g}_{k}^{d}=\sigma_{k\tilde{m_{k}}}^{2}$ as the direct-link
gain, and $\bar{g}_{k}^{c}=\underset{l\neq k}{\textrm{max}}\:\sigma_{k\tilde{m}_{l}}^{2}$
as the maximum cross-link gain. Define $\eta=\underset{k}{\textrm{min}\:}\bar{g}_{k}^{d}/\bar{g}_{k}^{c}$.
\begin{thm}
[Asymptotic Decoupling and Capacity Scaling]\label{thm:Asymptotic-Decoupling}For
any $\eta_{0}>0$, we have
\begin{equation}
\textrm{Pr}\left(\eta>\eta_{0}\right)\geq1-K\left[1-\left(1-\pi\left(\eta_{0}^{1/\zeta}+1\right)^{2}/\left(2M\right)\right)^{K-1}\right].\label{eq:ProDCgain}
\end{equation}
Furthermore, for any $\epsilon>0$, the maximum achievable sum-rate
$C_{s}$ almost surely satisfies 
\[
O\left(K\left(\frac{\zeta}{2}-\epsilon\right)\textrm{log}M\right)\leq C_{s}\leq O\left(K\left(\frac{\zeta}{2}+\epsilon\right)\textrm{log}M\right),
\]
 as $M\rightarrow\infty$ with $K,S$ fixed.\hfill \IEEEQED
\end{thm}

Please refer to Appendix \ref{sub:Sketch-ThmDecoupling} for the proof.
\begin{rem}
[Asymptotic Decoupling]For large $M/K$, there is a high probability
that $\eta$ is large, i.e., there is a high chance that each user
can find a set of nearby transmit antennas which are far from other
users. Due to this decoupling effect, simplified physical layer processing
(such as Matched-Filter precoder \cite{Marzetta_SPM12_LargeMIMO})
can also achieve good performance.\hfill \IEEEQED
\end{rem}

\begin{cor}
[Asymptotic Optimality of Algorithm S2]\label{cor:Asymptotic-Optimality-ofS3}
Algorithm S2 is asymptotically optimal, i.e., for any $\epsilon>0$,
the achieved sum-rate $\mathcal{I}_{\mathcal{A}}$ satisfies $\mathcal{I}_{\mathcal{A}}\overset{a.s}{\geq}O\left(K\left(\frac{\zeta}{2}-\epsilon\right)\textrm{log}M\right)$
as $M\rightarrow\infty$ with $K,S$ fixed.
\end{cor}

The proof is given in Appendix \ref{sub:Sketch-OptS3}.

\section{Numerical Results\label{sec:Numerical-Results}}

Consider a C-RAN serving $K$ users lying inside a square with an
area of $2\textrm{km}\times2\textrm{km}$. The simulation setup is
the same as that in Definition \ref{def:VLDAS} with path loss factor
$\zeta=2.5$.

\subsection{Accuracy of the Asymptotic Expressions}

\begin{figure}
\begin{centering}
\textsf{\includegraphics[clip,width=80mm]{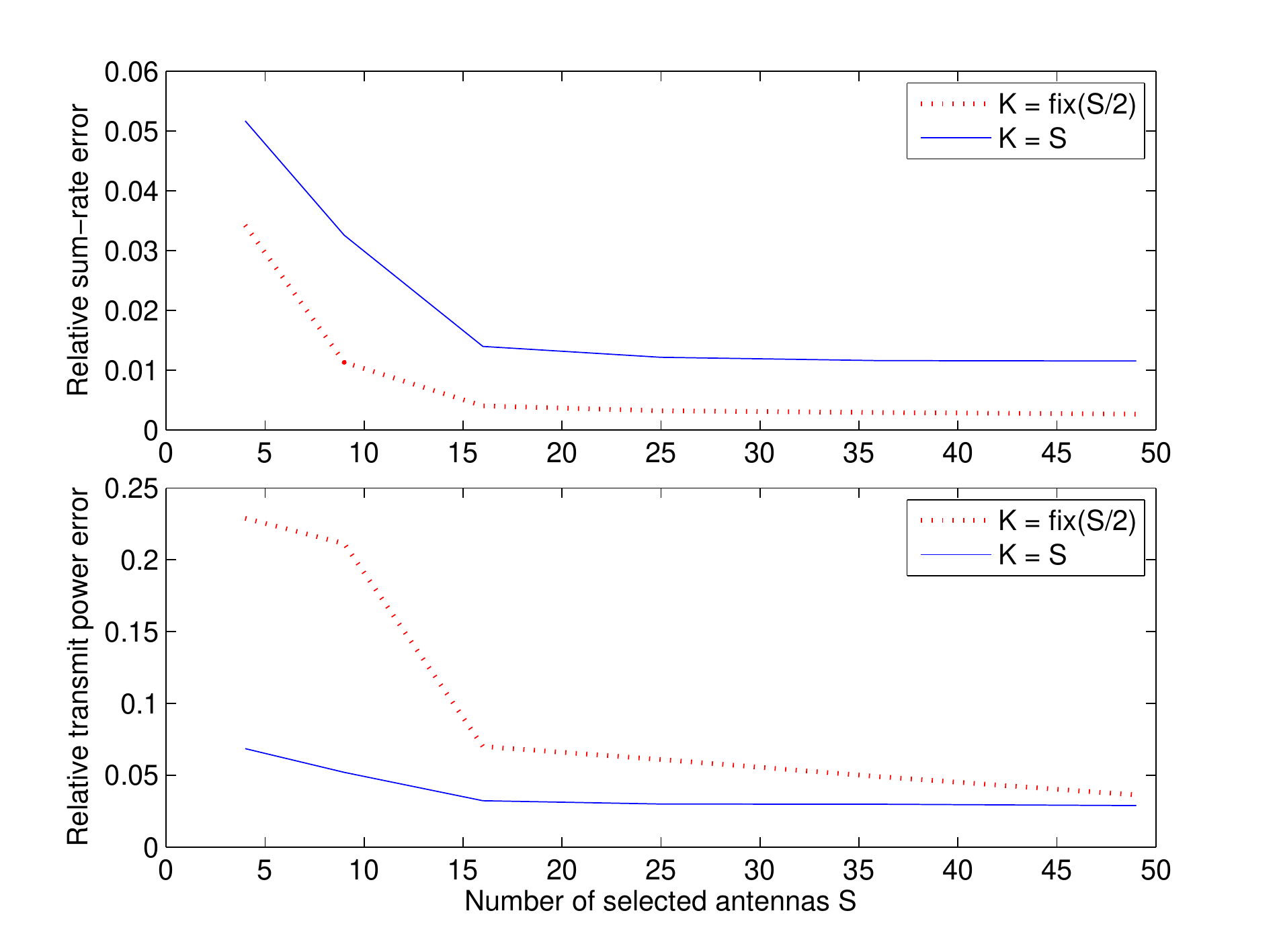}}
\par\end{centering}

\caption{\label{fig:Asym_err}Relative sum-rate/transmit power error versus
$S$}
\end{figure}

We verify the accuracy of the DE of sum-rate (i.e., $w_{k}=1,\:\forall k$)
$\bar{\mathcal{I}}\left(\mathcal{A},\alpha,\mathbf{p}\right)$ and
the DE of per antenna transmit power $\bar{P}_{m}\left(\mathcal{A},\alpha,\mathbf{p}\right),\forall m\in\mathcal{A}$
by comparing them to those obtained by Monte-Carlo simulations: $\mathcal{I}^{\textrm{sim}}$
and $P_{m}^{\textrm{sim}},\forall m\in\mathcal{A}$. We set $\mathcal{A}=\left\{ 1,...,M\right\} $,
i.e., $S=M$. Assume the per antenna power constraint is given by
$\rho_{m}=10\textrm{dB},\forall m$ and equal power allocation is
adopted, i.e., $\mathbf{P}=c\mathbf{I}$ in (\ref{eq:PrecodingM}),
where $c$ is chosen such that $\bar{P}_{m}\left(\mathcal{A},\alpha,\mathbf{p}\right)=\frac{10}{S}$.
The regularization factor is fixed as $\alpha=\beta/10$. In Fig.
\ref{fig:Asym_err}, we plot the relative sum-rate error $\left|\bar{\mathcal{I}}\left(\mathcal{A},\alpha,\mathbf{p}\right)-\mathcal{I}^{\textrm{sim}}\right|/\mathcal{I}^{\textrm{sim}}$
and the relative transmit power error $\left(\sum_{m\in\mathcal{A}}\left|\bar{P}_{m}\left(\mathcal{A},\alpha,\mathbf{p}\right)-P_{m}^{\textrm{sim}}\right|^{2}\right)^{1/2}/\left(\sum_{m\in\mathcal{A}}\left|P_{m}^{\textrm{sim}}\right|^{2}\right)^{1/2}$,
versus $S$. Both cases with $K=S/2$ (i.e., $\beta=1/2$) and $K=S$
(i.e., $\beta=1$) are simulated. The large scale fading matrix $\mathbf{\Sigma}$
is generated according to a random realization of user locations.
It can be seen that the asymptotic approximation is quite accurate.

In the rest of the simulations (i.e., in Fig. \ref{fig:Fig_BS_edge_even}
and \ref{fig:Fig2}), the per antenna power constraint is set as $\rho_{m}=P_{0},\forall m$.
The number of users $K$ is fixed as 8. From user 1 to user 8, the
weights $w_{k}$'s increase linearly from $0.5$ to $1.5$.

\subsection{Performance Gain of the Proposed Scheme w.r.t. Baseline}

\begin{figure}
\begin{centering}
\textsf{\includegraphics[clip,width=80mm]{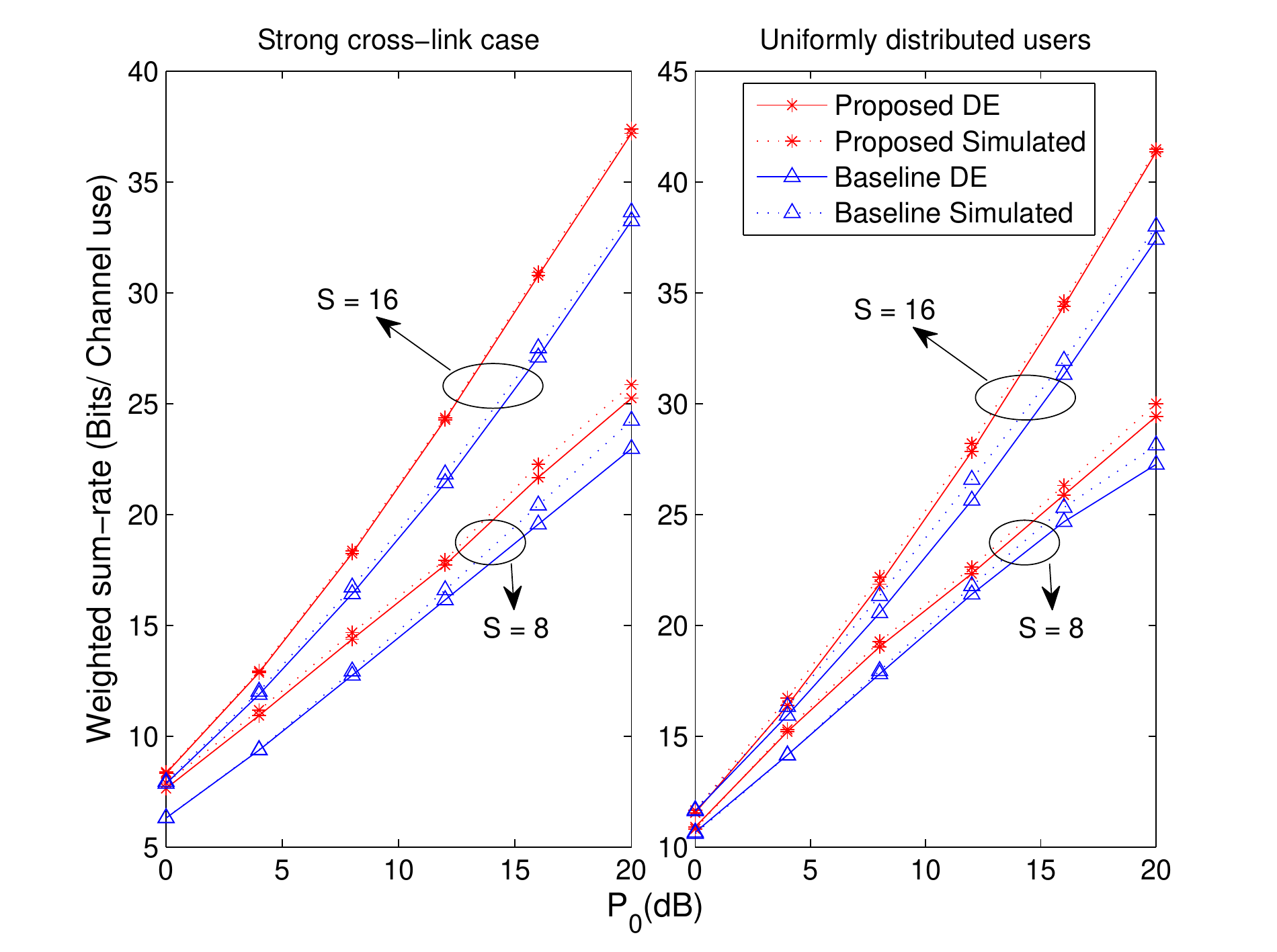}}
\par\end{centering}

\caption{\label{fig:Fig_BS_edge_even}Comparison of proposed antenna selection
scheme and baseline}
\end{figure}

In Fig. \ref{fig:Fig_BS_edge_even}, we verify the performance gain
of the proposed scheme w.r.t. the traditional antenna selection baseline
algorithm where each user is associated with the strongest antennas.
There are a total number of $M=25$ antennas. We plot the weighted
sum-rates averaged over different realizations of user locations versus
$P_{0}$ for $S=8$ and $16$ respectively. In the subplot on the
left-hand side of Fig. \ref{fig:Fig_BS_edge_even}, we consider the
strong cross link case, where the user locations are randomly generated
but with the restriction that the distance between each user and the
nearest antenna must be larger than a threshold. In this case, it
can be seen that the proposed scheme achieves significant performance
gain compared with the baseline. In the subplot on the right-hand
side of Fig. \ref{fig:Fig_BS_edge_even}, we consider the normal case
where the users are uniformly distributed. Similar results can be
observed, although the performance gain is smaller%
\footnote{This is because the unfavored scenarios for the baseline algorithm
as illustrated in Example \ref{exa:Strong-cross-link-causes} and
\ref{exa:Strong-cross-link-provides} occur less frequently when the
users are uniformly distributed.%
}.

\subsection{Advantages of the Proposed Scheme over the Cases without Antenna
Selection}

\begin{figure}
\begin{centering}
\textsf{\includegraphics[clip,width=85mm]{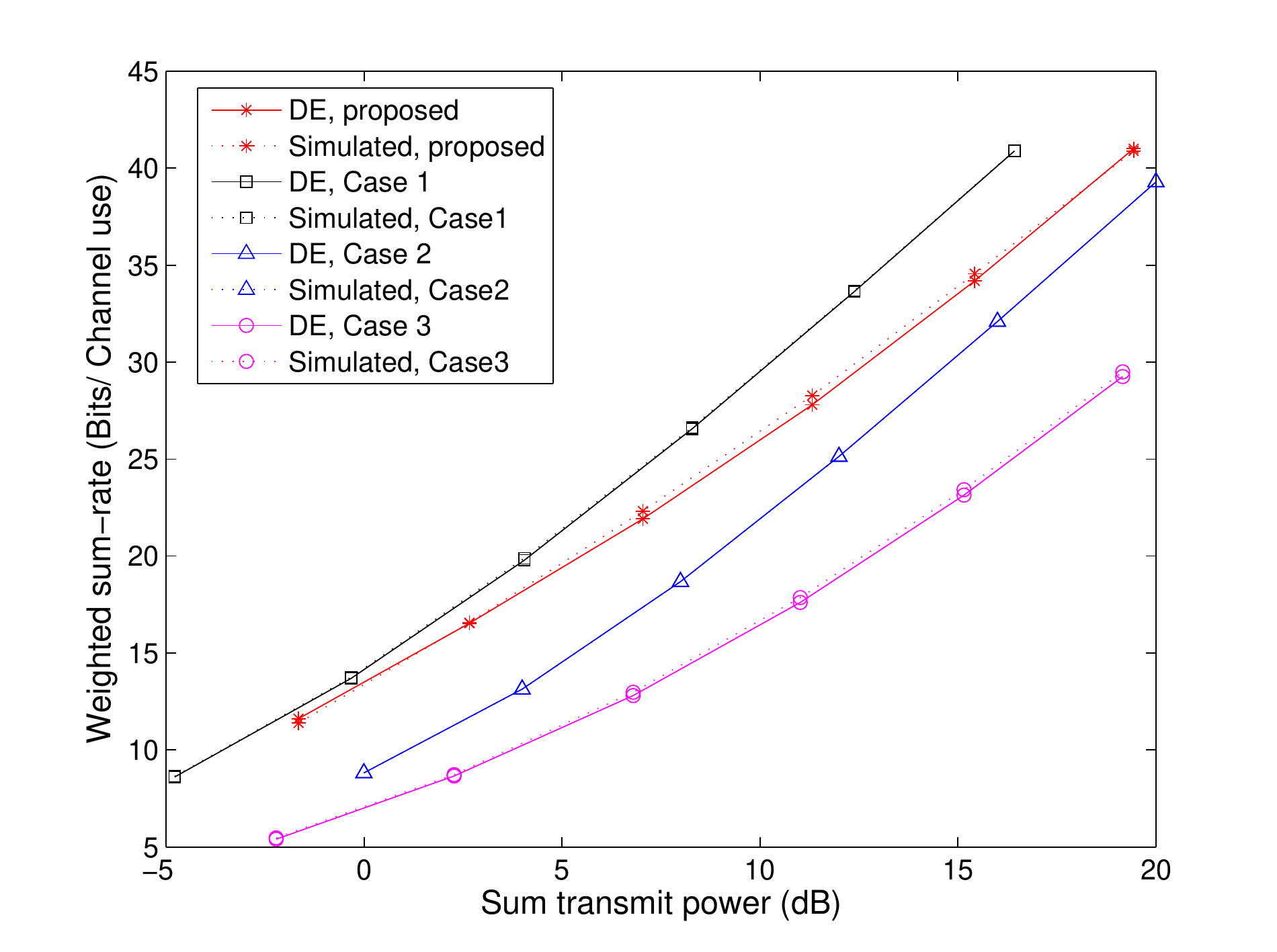}}
\par\end{centering}

\caption{\label{fig:Fig2}Comparison of weighted sum-rates for different choices
of transmit antennas}
\end{figure}

In Fig. \ref{fig:Fig2}, we compare the proposed antenna selection
scheme with various cases without antenna selection. There are a total
number of $M=49$ antennas and $S=16$ of them are selected for transmission.
The performance of the following cases are compared. Case 1: All the
$49$ distributed antennas are used for transmission. Case 2: All
the $49$ antennas are collocated at the BS and are used for transmission.
Case 3: There is a total of $16$ antennas evenly distributed in the
square and all the 16 antennas are used for transmission. We plot
the weighted sum-rates averaged over different realizations of user
locations versus the sum transmit power%
\footnote{In Fig. \ref{fig:Fig2}, the horizontal axis is chosen to be sum transmit
power to make a fair comparison of the performance for the cases with
different number of active transmit antennas and different antenna
deployments (i.e., distributed versus collocated).%
}. The following advantages of the proposed antenna selection scheme
can be observed. 1) Under the same sum transmit power, it achieves
a weighted sum-rate that is close to Case 1, and is better than Case
2, while the pilot training overhead is lower. 2) The performance
is much better than Case 3 due to large antenna gain.

\section{Conclusion\label{sec:Conlusion}}

In this paper, we have considered a downlink antenna selection in
a large distributed MIMO network with $M\gg1$ distributed antennas
serving $K$ users using RZF precoding. The objective is to maximize
the average weighted sum-rate under per antenna and sum power constraints
based on large scale fading. This mixed combinatorial and non-convex
problem is decomposed into simpler subproblems, each of which is then
solved by an efficient algorithm. We also show that the capacity of
a very large distributed MIMO network scales according to $O\left(K\frac{\zeta}{2}\textrm{log}M\right)$,
where $\zeta$ is the path loss factor.

\appendix

\subsection{Proof of Proposition \ref{prop:boundSpnorm}\label{sub:Proof-of-PropositionSPbound}}

First, we prove the following lemma.
\begin{lem}
\label{lem:Spbound}Define $\sigma_{\textrm{max}}\triangleq\underset{1\leq k\leq K,1\leq m\leq M}{\textrm{max}}\sigma_{km}$.
Then for every $t\geq0$, with probability at least $1-e^{-t^{2}/\sigma_{\textrm{max}}^{2}}$
one has
\begin{equation}
\left\Vert \mathbf{H}\right\Vert \leq C\left(\sigma_{\textrm{max}}\right)\left(\sqrt{K}+\sqrt{M}\right)+t,\label{eq:SpboundProb}
\end{equation}
where $C\left(\sigma_{\textrm{max}}\right)$ is a constant only depends
on $\sigma_{\textrm{max}}$.\end{lem}
\begin{IEEEproof}
Recall that $\mathbf{H}=\mathbf{W}\circ\mathbf{\Sigma}$, where the
notation $\circ$ denotes the Hadamard product; and $\mathbf{W}$
is the small scale fading matrix with i.i.d.$\mathcal{\sim CN}\left(\mathbf{0},1\right)$
entries. For any matrix $\mathbf{X}\in\mathbb{C}^{K\times M}$, let
$\mathbf{x}=\textrm{Vec}\left(\mathbf{X}\right)$ represent the vector
obtained by stacking the column of $\mathbf{X}$, and let $\textrm{Vec}^{-1}\left(\mathbf{x}\right)$
denote the inverse mapping. Define a function $f\left(\mathbf{x}\right)\triangleq\left\Vert \textrm{Vec}{}^{-1}\left(\mathbf{x}\right)\circ\mathbf{\Sigma}\right\Vert $.
Then we have 
\begin{eqnarray}
\left|f\left(\mathbf{x}\right)-f\left(\mathbf{y}\right)\right| & \leq & \left\Vert \mathbf{x}\circ\textrm{Vec}\left(\mathbf{\Sigma}\right)-\mathbf{y}\circ\textrm{Vec}\left(\mathbf{\Sigma}\right)\right\Vert \label{eq:LNotm}\\
 & \leq & \sigma_{\textrm{max}}\left\Vert \mathbf{x}-\mathbf{y}\right\Vert ,\nonumber 
\end{eqnarray}
where (\ref{eq:LNotm}) follows from the fact that $\left\Vert \mathbf{x}\circ\textrm{Vec}\left(\mathbf{\Sigma}\right)\right\Vert $
is a 1-Lipschitz function of $\mathbf{x}\circ\textrm{Vec}\left(\mathbf{\Sigma}\right)$.
It can be deduced from \cite[Theorem 2]{Latala_PAM05_spnormbound}
that
\begin{equation}
\textrm{E}\left\Vert \mathbf{H}\right\Vert \leq C\left(\sigma_{\textrm{max}}\right)\left(\sqrt{K}+\sqrt{M}\right).\label{eq:Ebound}
\end{equation}
Note that $\left\Vert \mathbf{H}\right\Vert =f\left(\mathbf{w}\right)$,
where $\mathbf{w}=\textrm{Vec}\left(\mathbf{W}\right)$ has i.i.d.$\mathcal{\sim CN}\left(\mathbf{0},1\right)$
entries. One has
\begin{eqnarray}
 &  & \textrm{Pr}\left\{ f\left(\mathbf{w}\right)-C\left(\sigma_{\textrm{max}}\right)\left(\sqrt{K}+\sqrt{M}\right)>t\right\} \nonumber \\
 & \leq & \textrm{Pr}\left\{ f\left(\mathbf{w}\right)-\textrm{E}f\left(\mathbf{w}\right)>t\right\} \leq e^{-t^{2}/\sigma_{\textrm{max}}^{2}},\label{eq:GaussCon}
\end{eqnarray}
where the last inequality is obtained by applying the Gaussian concentration
in \cite[Prop. 5.34]{Vershynin_CUP12_RMT} on the function $\hat{f}\left(\hat{\mathbf{w}}\right)\triangleq f\left(\mathbf{w}\right),$
where $\hat{\mathbf{w}}\triangleq\left[\begin{array}{cc}
\textrm{Re}(\sqrt{2}\mathbf{w}) & \textrm{Im}(\sqrt{2}\mathbf{w})\end{array}\right]^{T}$. This completes the proof for Lemma \ref{lem:Spbound}.
\end{IEEEproof}

Noting that $\bar{\mathbf{H}}=\mathbf{H}/\sqrt{S}$, Proposition \ref{prop:boundSpnorm}
follows immediately from Lemma \ref{lem:Spbound} by setting $t=\sqrt{S}$
and letting $S\rightarrow\infty$.

\subsection{Proof of Lemma \ref{lem:AsyPm}\label{sub:Proof-of-mainTheorem}}

For conciseness, the $\mathbf{\Sigma}^{(S)},\Psi^{(S)}\left(\mathbf{\Sigma}^{(S)}\right)=\left\{ \mathcal{A}^{(S)},\alpha^{(S)},\mathbf{p}^{(S)}\right\} $
are denoted as $\mathbf{\Sigma},\Psi\left(\mathbf{\Sigma}\right)=\left\{ \mathcal{A},\alpha,\mathbf{p}\right\} $
and we use ``$\overset{a.s}{\rightarrow}0$'' as a simplified notation
for ``$\overset{a.s}{\rightarrow}0,\:\textrm{as}\: S\rightarrow\infty$''.
Recall that $\mathcal{A}_{j}$ is the $j^{\textrm{th}}$ antenna in
$\mathcal{A}$. By denoting $\tilde{\mathbf{Q}}_{j}=\bar{\mathbf{H}}_{j}^{\textrm{c}}\bar{\mathbf{H}}_{j}^{\textrm{c}\dagger}+\alpha\mathbf{I}_{K}$,
where $\bar{\mathbf{H}}_{j}^{\textrm{c}}$ is the matrix of $\bar{\mathbf{H}}$
where the $j^{\textrm{th}}$ column is removed, and applying matrix
inverse lemma to (\ref{eq:Pm}), we have
\begin{equation}
SP_{\mathcal{A}_{j}}\left(\Psi\left(\mathbf{\Sigma}\right)\right)=A_{j}^{c}\left(1+\bar{\mathbf{h}}_{j}^{c\dagger}\tilde{\mathbf{Q}}_{j}^{-1}\bar{\mathbf{h}}_{j}^{c}\right)^{-2},\label{eqSPA}
\end{equation}
where $A_{j}^{c}=\bar{\mathbf{h}}_{j}^{c\dagger}\tilde{\mathbf{Q}}_{j}^{-1}\mathbf{P}\tilde{\mathbf{Q}}_{j}^{-1}\bar{\mathbf{h}}_{j}^{c}$,
and $\bar{\mathbf{h}}_{j}^{c}=\bar{\mathbf{H}}\mathbf{1}_{j}$ is
the $j$-th column of $\bar{\mathbf{H}}$. Note that
\begin{equation}
A_{j}^{c}=\tilde{\mathbf{h}}_{j}^{c\dagger}\boldsymbol{\sigma}_{j}^{c}\tilde{\mathbf{Q}}_{j}^{-1}\mathbf{P}\tilde{\mathbf{Q}}_{j}^{-1}\boldsymbol{\sigma}_{j}^{c}\tilde{\mathbf{h}}_{j}^{c},\label{eq:Akc}
\end{equation}
where $\boldsymbol{\sigma}_{j}^{c}=\textrm{diag}\left(\sigma_{1j},...,\sigma_{Kj}\right),$
and the elements of $\tilde{\mathbf{h}}_{j}^{c}\triangleq\left(\boldsymbol{\sigma}_{j}^{c}\right)^{-1}\bar{\mathbf{h}}_{j}^{c}$
are i.i.d. complex random variables with zero mean, variance $1/S$.
For any $\mathbf{A}\in\mathbb{C}^{K\times K}$, define $\Xi_{\boldsymbol{\sigma}_{j}^{c},\mathbf{A}}\triangleq\frac{1}{K}\textrm{Tr}\left(\left(\boldsymbol{\sigma}_{j}^{c}\right)^{2}\left(\mathbf{A}+\alpha\mathbf{I}_{K}\right)^{-1}\right)$.
Using \cite[Corrolary 1]{Tse_TIT00_LSP}, (\ref{eq:Akc}) and the
equality $\frac{1}{K}\textrm{Tr}\left(\left(\boldsymbol{\sigma}_{j}^{c}\right)^{2}\tilde{\mathbf{Q}}_{j}^{-1}\mathbf{P}\tilde{\mathbf{Q}}_{j}^{-1}\right)=-\left.\frac{\partial}{\partial z}\Xi_{\boldsymbol{\sigma}_{j}^{c},\bar{\mathbf{H}}_{j}^{\textrm{c}}\bar{\mathbf{H}}_{j}^{\textrm{c}\dagger}+z\mathbf{P}}\right|_{z=0}$,
we have 
\begin{equation}
A_{j}^{c}+\frac{K}{S}\left.\frac{\partial}{\partial z}\Xi_{\boldsymbol{\sigma}_{j}^{c},\bar{\mathbf{H}}_{j}^{\textrm{c}}\bar{\mathbf{H}}_{j}^{\textrm{c}\dagger}+z\mathbf{P}}\right|_{z=0}\overset{a.s}{\rightarrow}0.\label{eq:Akc-1}
\end{equation}
By \cite[Lemma 2.1]{Bai_AAP07_TRRMT}, we have 
\begin{equation}
\left.\frac{\partial}{\partial z}\Xi_{\boldsymbol{\sigma}_{j}^{c},\bar{\mathbf{H}}_{j}^{\textrm{c}}\bar{\mathbf{H}}_{j}^{\textrm{c}\dagger}+z\mathbf{P}}\right|_{z=0}-\left.\frac{\partial}{\partial z}\Xi_{\boldsymbol{\sigma}_{j}^{c},\bar{\mathbf{H}}\bar{\mathbf{H}}^{\dagger}+z\mathbf{P}}\right|_{z=0}\overset{a.s}{\rightarrow}0.\label{eq:Trbk2}
\end{equation}
Applying \cite[Theorem 1]{Wagner_TIT12s_LargeMIMO} to $\Xi_{\boldsymbol{\sigma}_{j}^{c},\bar{\mathbf{H}}\bar{\mathbf{H}}^{\dagger}+z\mathbf{P}}$,
we obtain%
\footnote{In \cite[Thereom 1]{Wagner_TIT12s_LargeMIMO}, $\alpha$ is a positive
constant. However, it can be verified that the proof of \cite[Thereom 1]{Wagner_TIT12s_LargeMIMO}
still holds if we replace the constant $\alpha$ by $\alpha^{(S)}=\Psi_{\alpha}^{(S)}\left(\mathbf{\Sigma}^{(S)}\right)$
due to the bounded constraint on $\Psi_{\alpha}^{(S)}$.%
}
\begin{equation}
\frac{K}{S}\left.\frac{\partial}{\partial z}\Xi_{\boldsymbol{\sigma}_{j}^{c},\bar{\mathbf{H}}\bar{\mathbf{H}}^{\dagger}+z\mathbf{P}}\right|_{z=0}+\frac{\alpha^{-1}}{S}\sum_{i=1}^{K}\sigma_{i\mathcal{A}_{j}}^{2}\left(p_{i}v_{i}-\varphi_{i}\right)\overset{a.s}{\rightarrow}0,\label{eq:Trbk3}
\end{equation}
where $v_{i}$ is defined in (\ref{eq:vfix}) and $\varphi_{i}$ is
defined in (\ref{eq:FaiDef}). Combining (\ref{eq:Akc-1}-\ref{eq:Trbk3}),
we have 
\begin{equation}
A_{j}^{c}-\frac{\alpha^{-1}}{S}\sum_{i=1}^{K}\sigma_{i\mathcal{A}_{j}}^{2}\left(p_{i}v_{i}-\varphi_{i}\right)\overset{a.s}{\rightarrow}0.\label{eq:aKC2}
\end{equation}

Note that $\left(\alpha\mathbf{I}_{K}+\mathbf{\Delta}-\mathbf{C}\right)^{-1}$
in (\ref{eq:FaiDef}) is invertible because it is diagonally dominant.
Applying \cite[Corrolary 1]{Tse_TIT00_LSP}, \cite[Lemma 2.1]{Bai_AAP07_TRRMT}
and \cite[Theorem 1]{Wagner_TIT12s_LargeMIMO} one by one, we have
$\bar{\mathbf{h}}_{j}^{c\dagger}\tilde{\mathbf{Q}}_{j}^{-1}\bar{\mathbf{h}}_{j}^{c}-\frac{K}{S}\frac{1}{K}\textrm{Tr}\left(\left(\boldsymbol{\sigma}_{j}^{c}\right)^{2}\tilde{\mathbf{Q}}_{j}^{-1}\right)\overset{a.s}{\rightarrow}0$,
$\frac{1}{K}\textrm{Tr}\left(\left(\boldsymbol{\sigma}_{j}^{c}\right)^{2}\tilde{\mathbf{Q}}_{j}^{-1}\right)-\frac{1}{K}\textrm{Tr}\left(\left(\boldsymbol{\sigma}_{j}^{c}\right)^{2}\mathbf{Q}^{-1}\right)\overset{a.s}{\rightarrow}0$
and $\frac{1}{K}\textrm{Tr}\left(\left(\boldsymbol{\sigma}_{j}^{c}\right)^{2}\mathbf{Q}^{-1}\right)-\frac{1}{K}\sum_{i=1}^{K}\sigma_{i\mathcal{A}_{j}}^{2}v_{i}\overset{a.s}{\rightarrow}0$,
where $\tilde{\mathbf{Q}}=\bar{\mathbf{H}}\bar{\mathbf{H}}^{\dagger}+\alpha\mathbf{I}_{K}$.
Hence, 
\begin{equation}
\bar{\mathbf{h}}_{j}^{c\dagger}\tilde{\mathbf{Q}}_{j}^{-1}\bar{\mathbf{h}}_{j}^{c}-\frac{1}{S}\sum_{i=1}^{K}\sigma_{i\mathcal{A}_{j}}^{2}v_{i}\overset{a.s}{\rightarrow}0.\label{eq:HqH}
\end{equation}
Combining (\ref{eqSPA}), (\ref{eq:aKC2}) and (\ref{eq:HqH}), we
have $SP_{m}\left(\Psi\left(\mathbf{\Sigma}\right)\right)-S\bar{P}_{m}\left(\Psi\left(\mathbf{\Sigma}\right)\right)\overset{a.s}{\rightarrow}0,\forall m\in\mathcal{A}$.

\subsection{Proof of Theorem \ref{thm:AsyEqP} \label{sub:Proof-of-Theorem_AsyEqP}}

Let $\Psi^{(S)\circ}\left(\mathbf{\Sigma}^{(S)}\right)$ be the optimal
solution of Problem $\mathcal{P}\left(\mathbf{\Sigma}^{(S)}\right)$.
For an admissible policy $\Psi^{(S)}$, let $\left\{ \mathcal{A}^{(S)},\alpha^{(S)},\mathbf{p}^{(S)}\right\} =\Psi^{(S)}\left(\mathbf{\Sigma}^{(S)}\right)$
and define $Y{}_{m}^{\left(S\right)}\triangleq SP_{m}\left(\Psi^{(S)}\left(\mathbf{\Sigma}^{(S)}\right)\right)$
for any $m\in\mathcal{A}^{(S)}$. Recall that $K=\left\lceil \beta S\right\rceil $
and $M=\left\lceil \overline{\beta}S\right\rceil $. Then according
to the definition of admissible policy and Assumption \ref{asm:LSFM},
we have$\underset{1\leq k\leq K}{\textrm{sup}}p_{k}^{(S)}\leq P_{\textrm{max}}$,
$\alpha^{(S)}\geq\alpha_{\textrm{min}}$ and $\underset{1\leq k\leq K,1\leq m\leq M}{\textrm{sup}}\sigma_{km}^{(S)}\leq\sigma_{\textrm{max}}$
for some $\sigma_{\textrm{max}}>0$, uniformly on $S$. Using the
expression of the per antenna transmit power in (\ref{eq:Pm}), it
can be shown that $Y{}_{m}^{\left(S\right)}\leq\frac{\left(\beta+1\right)P_{\textrm{max}}\sigma_{\textrm{max}}^{2}}{\alpha_{\textrm{min}}^{2}}\overline{Y}{}_{m}^{\left(S\right)}$,
where $\overline{Y}{}_{m}^{\left(S\right)}\triangleq\frac{\sum_{k=1}^{K}\left|W_{km}\right|^{2}}{K}$
and $W_{km}\mathcal{\sim CN}\left(\mathbf{0},1\right)$ is the element
at the $k$-th row and $m$-th column of the small scale fading matrix
$\mathbf{W}$. Note that $2K\overline{Y}{}_{m}^{\left(S\right)}\sim\chi^{2}\left(2K\right)$
is a chi-square random variable with $2K$ degrees of freedom. Using
the Chernoff bounds on the upper tails of the CDF of $\chi^{2}\left(2K\right)$,
we have $F_{\overline{Y}}^{(S)}\left(t\right)\triangleq\Pr\left[\overline{Y}{}_{m}^{\left(S\right)}\geq t\right]\leq\left(te^{1-t}\right)^{K}\leq te^{1-t}$
for any $t\geq1$. Using the relationship between expectation and
CDF for a positive random variable, it can be shown that $\textrm{E}\left[\overline{Y}{}_{m}^{\left(S\right)}I_{\overline{Y}{}_{m}^{\left(S\right)}\geq Y}\right]=\int_{Y}^{\infty}F_{\overline{Y}}^{(S)}\left(t\right)dt+YF_{\overline{Y}}^{(S)}\left(Y\right)\leq\left(1+Y+Y^{2}\right)e^{1-Y}$
for any $Y\geq1$, where $I_{\overline{Y}{}_{m}^{\left(S\right)}\geq Y}=1$
if $\overline{Y}{}_{m}^{\left(S\right)}\geq Y$ and $I_{\overline{Y}{}_{m}^{\left(S\right)}\geq Y}=0$
otherwise. It follows that $\left\{ \overline{Y}{}_{m}^{\left(S\right)}\right\} $
is uniformly integrable \cite{David_CUP97_Prob}. Hence, $\left\{ Y{}_{m}^{\left(S\right)}\right\} $
is also uniformly integrable. Combine the above and Lemma \ref{lem:AsyPm},
it follows that 
\begin{equation}
S\textrm{E}\left[P_{m}\left(\Psi^{*}\left(\mathbf{\Sigma}\right)\right)\left|\mathbf{\Sigma}\right.\right]-S\bar{P}_{m}\left(\Psi^{*}\left(\mathbf{\Sigma}\right)\right)\rightarrow0.\label{eq:Pgap0A}
\end{equation}
Note that for conciseness, $\mathbf{\Sigma}^{(S)},\Psi^{(S)\circ}\left(\mathbf{\Sigma}^{(S)}\right),\Psi^{(S)*}\left(\mathbf{\Sigma}^{(S)}\right)$
are denoted as $\mathbf{\Sigma},\Psi^{\circ}\left(\mathbf{\Sigma}\right),\Psi^{*}\left(\mathbf{\Sigma}\right)$
when there is no ambiguity and we use ``$\rightarrow0$'' as a simplified
notation for ``$\rightarrow0$ $\textrm{as}\: S\rightarrow\infty$''.
By the definition of $\bar{P}_{m}\left(\Psi^{*}\left(\mathbf{\Sigma}\right)\right)$,
we have
\begin{eqnarray}
S\bar{P}_{m}\left(\Psi^{*}\left(\mathbf{\Sigma}\right)\right)-\rho_{m} & \leq & 0.\label{eq:Pgap0}
\end{eqnarray}
Then it follows from (\ref{eq:Pgap0A}) and (\ref{eq:Pgap0}) that,
as $S\rightarrow\infty$,
\begin{equation}
S\textrm{E}\left[P_{m}\left(\Psi^{*}\left(\mathbf{\Sigma}\right)\right)\left|\mathbf{\Sigma}\right.\right]-\rho_{m}\leq0,\forall m\in\Psi_{\mathcal{A}}^{*}\left(\mathbf{\Sigma}\right).\label{eq:Pgapf1}
\end{equation}
Similarly, as $S\rightarrow\infty,$ it can be shown that 
\begin{equation}
S\bar{P}_{m}\left(\Psi^{\circ}\left(\mathbf{\Sigma}\right)\right)-\rho_{m}\leq0,\:\forall m\in\Psi_{\mathcal{A}}^{\circ}\left(\mathbf{\Sigma}\right).\label{eq:Pgapf2}
\end{equation}

Similarly, define $Z_{k}^{(S)}\triangleq\textrm{log}\left(1+\gamma_{k}\left(\Psi^{(S)}\left(\mathbf{\Sigma}^{(S)}\right)\right)\right)$
and $\varepsilon_{k}^{\left(S\right)}\triangleq\textrm{log}\left(1+\bar{\gamma}_{k}\left(\Psi^{(S)}\left(\mathbf{\Sigma}^{(S)}\right)\right)\right)-\textrm{E}\left[Z_{k}^{(S)}\left|\mathbf{\Sigma}^{(S)}\right.\right]$.
Since $Kw_{k}^{(S)}$ is uniformly bounded on $S$ according to Assumption
\ref{asm:LSFM}, there exists a constant $C>0$ such that $\underset{1\leq k\leq K}{\textrm{sup}}w_{k}^{(S)}\leq\frac{C}{K}$,
uniformly on $S$. Using the expression of the SINR in (\ref{eq:SINR}),
it can be shown that $Z_{k}^{(S)}\leq\textrm{log}\left(1+P_{\textrm{max}}\right)$.
Hence, $\left\{ Z_{k}^{(S)}\right\} $ is uniformly integrable \cite{David_CUP97_Prob}.
Together with Lemma \ref{lem:AsyWS}, it follows that $\varepsilon_{k}^{\left(S\right)}\rightarrow0$,
i.e., $\forall\delta>0$, $\exists S_{0}>0$ such that $\forall S\geq S_{0}$,
$\left|\varepsilon_{k}^{\left(S\right)}\right|<\frac{\delta}{C},\forall k$.
Then, $\forall\delta>0$, $\exists S_{0}>0$ such that $\forall S\geq S_{0}$,
$\left|\sum_{k=1}^{K}w_{k}^{(S)}\varepsilon_{k}^{\left(S\right)}\right|\leq\sum_{k=1}^{K}w_{k}^{(S)}\left|\varepsilon_{k}^{\left(S\right)}\right|\leq\frac{C}{K}\sum_{k=1}^{K}\left|\varepsilon_{k}^{\left(S\right)}\right|<\delta$,
i.e., $\sum_{k=1}^{K}w_{k}^{(S)}\varepsilon_{k}^{\left(S\right)}\rightarrow0$.
Note that $\sum_{k=1}^{K}w_{k}^{(S)}\varepsilon_{k}^{\left(S\right)}=\mathcal{I}\left(\Psi^{(S)}\left(\mathbf{\Sigma}^{(S)}\right)|\mathbf{\Sigma}^{(S)}\right)-\bar{\mathcal{I}}\left(\Psi^{(S)}\left(\mathbf{\Sigma}^{(S)}\right)|\mathbf{\Sigma}^{(S)}\right)$.
Then it follows from the above analysis that 
\begin{eqnarray}
\mathcal{I}\left(\Psi^{\circ}\left(\mathbf{\Sigma}\right)|\mathbf{\Sigma}\right)-\bar{\mathcal{I}}\left(\Psi^{\circ}\left(\mathbf{\Sigma}\right)|\mathbf{\Sigma}\right) & \rightarrow & 0,\nonumber \\
\bar{\mathcal{I}}\left(\Psi^{*}\left(\mathbf{\Sigma}\right)|\mathbf{\Sigma}\right)-\mathcal{I}\left(\Psi^{*}\left(\mathbf{\Sigma}\right)|\mathbf{\Sigma}\right) & \rightarrow & 0.\label{eq:gapI0}
\end{eqnarray}
From (\ref{eq:Pgapf1}), (\ref{eq:Pgapf2}) and the definition of
$\Psi^{*}\left(\mathbf{\Sigma}\right)$ and $\Psi^{\circ}\left(\mathbf{\Sigma}\right)$,
we have
\begin{eqnarray}
\mathcal{I}\left(\Psi^{\circ}\left(\mathbf{\Sigma}\right)|\mathbf{\Sigma}\right)-\mathcal{I}\left(\Psi^{*}\left(\mathbf{\Sigma}\right)|\mathbf{\Sigma}\right) & \geq & 0,\nonumber \\
\bar{\mathcal{I}}\left(\Psi^{\circ}\left(\mathbf{\Sigma}\right)|\mathbf{\Sigma}\right)-\bar{\mathcal{I}}\left(\Psi^{*}\left(\mathbf{\Sigma}\right)|\mathbf{\Sigma}\right) & \leq & 0.\label{eq:Iequ}
\end{eqnarray}
It follows from (\ref{eq:gapI0}) and (\ref{eq:Iequ}) that
\[
\mathcal{I}\left(\Psi^{\circ}\left(\mathbf{\Sigma}\right)|\mathbf{\Sigma}\right)-\mathcal{I}\left(\Psi^{*}\left(\mathbf{\Sigma}\right)|\mathbf{\Sigma}\right)\rightarrow0.
\]
This completes the proof for Theorem \ref{thm:AsyEqP}.

\subsection{Proof of Proposition \ref{prop:optpboundr} \label{sub:Proof-of-Proposition}}

Using the notations defined in Section \ref{sub:Algorithm-S1-for},
the constraint in (\ref{eq:AsyCons}) can be rewritten as $\mathbf{R}\mathbf{p}\leq\boldsymbol{\rho}$.
It can be verified that $R_{j,l}>0,\forall j,l$, where $R_{j,l}$
is the element at the $j$-th row and $l$-th column of $\mathbf{R}$.
Suppose that there exists $k$ such that $p_{k}>P_{\textrm{max}}$.
We must have $\sum_{l=1}^{K}R_{j,l}p_{l}>R_{j,k}P_{\textrm{max}}>\rho_{\mathcal{A}_{j}}$
for sufficiently large $P_{\textrm{max}}$. Hence, for sufficiently
large $P_{\textrm{max}}$, we must have $\underset{1\leq k\leq K}{\textrm{max}}\: p_{k}\leq P_{\textrm{max}}$
in order to satisfy the per antenna power constraint in (\ref{eq:AsyCons}).
This completes the proof.

\subsection{Calculation of the Derivative $\frac{\partial\hat{\mathcal{I}}\left(\mathcal{A},\alpha\right)}{\partial\alpha}$\label{sub:Calculation-of-DIdruo}}

For convenience, define two $\left(S+K\right)$-dimensional vectors
\[
\boldsymbol{\rho}_{\textrm{ext}}=\left[\begin{array}{c}
\boldsymbol{\rho}\\
\mathbf{0}
\end{array}\right],\:\tilde{\boldsymbol{\lambda}}_{\textrm{ext}}=\left[\begin{array}{c}
\tilde{\boldsymbol{\lambda}}\\
-\tilde{\boldsymbol{\nu}}
\end{array}\right].
\]
Define a $\left(S+K\right)\times K$ matrix $\mathbf{R}_{\textrm{ext}}\triangleq\left[\mathbf{R}^{T},-\mathbf{I}_{K}\right]^{T}$.
Define a vector $\mathbf{e}\in\mathbb{R}^{K}$ whose $k^{\textrm{th}}$
element is
\begin{eqnarray*}
e_{k} & = & \sum_{l=1}^{K}\left(\frac{w_{l}g_{lk}\sum_{i=1}^{K}\tilde{p}_{i}\left(\mathcal{A},\alpha\right)\frac{\partial g_{li}}{\partial\alpha}}{\left(g_{ll}\tilde{p}_{l}\left(\mathcal{A},\alpha\right)+\tilde{\Omega}_{l}\right)^{2}}-\frac{w_{l}\frac{\partial g_{lk}}{\partial\alpha}}{g_{ll}\tilde{p}_{l}\left(\mathcal{A},\alpha\right)+\tilde{\Omega}_{l}}\right)\\
 &  & +\sum_{l\neq k}^{K}\left(\frac{w_{l}\frac{\partial g_{lk}}{\partial\alpha}}{\tilde{\Omega}_{l}}-\frac{w_{l}g_{lk}\sum_{i\neq l}^{K}\tilde{p}_{i}\left(\mathcal{A},\alpha\right)\frac{\partial g_{li}}{\partial\alpha}}{\tilde{\Omega}_{l}^{2}}\right).
\end{eqnarray*}
Define a $K\times K$ matrix $\mathbf{\Upsilon}$ whose element at
the $k^{\textrm{th}}$ row and $l^{\textrm{th}}$ column is
\[
\Upsilon_{kl}=\sum_{l=1}^{K}\frac{-w_{l}g_{lk}g_{ln}}{\left(1+\sum_{i=1}^{K}g_{li}\tilde{p}_{i}\left(\mathcal{A},\alpha\right)\right)^{2}}+\sum_{l\neq k,n}\frac{w_{l}g_{lk}g_{ln}}{\tilde{\Omega}_{l}^{2}}.
\]
Finally, define a $\left(2K+S\right)\times\left(2K+S\right)$ matrix
\[
\mathbf{\Upsilon}_{\textrm{ext}}=\left[\begin{array}{cc}
\mathbf{\Upsilon}; & -\mathbf{R}_{\textrm{ext}}^{T}\\
\textrm{diag}\left(\tilde{\boldsymbol{\lambda}}_{\textrm{ext}}\right)\mathbf{R}_{\textrm{ext}}; & \textrm{diag}\left(\mathbf{R}_{\textrm{ext}}\tilde{\mathbf{p}}\left(\mathcal{A},\alpha\right)-\boldsymbol{\rho}_{\textrm{ext}}\right)
\end{array}\right].
\]

Taking partial derivative of the equations in (\ref{eq:KKT}) with
respect to $\alpha$, we obtain the following linear equations
\[
\mathbf{\Upsilon}_{\textrm{ext}}\left[\begin{array}{c}
\frac{\partial\tilde{\mathbf{p}}\left(\mathcal{A},\alpha\right)}{\partial\alpha}\\
\frac{\partial\tilde{\boldsymbol{\lambda}}_{\textrm{ext}}}{\partial\alpha}
\end{array}\right]=\left[\begin{array}{c}
\left(\frac{\partial\mathbf{R}}{\partial\alpha}\right)^{T}\tilde{\boldsymbol{\lambda}}+\mathbf{e}\\
-\textrm{diag}\left(\tilde{\boldsymbol{\lambda}}_{\textrm{ext}}\right)\left(\frac{\partial\mathbf{R}_{\textrm{ext}}}{\partial\alpha}\right)^{T}\tilde{\mathbf{p}}\left(\mathcal{A},\alpha\right)
\end{array}\right].
\]
Then we can obtain $\frac{\partial\tilde{\mathbf{p}}\left(\mathcal{A},\alpha\right)}{\partial\alpha}$
by solving the above linear equations. 

Define $\mathcal{J}=\left\{ j:\:\bar{P}_{\mathcal{A}_{j}}\left(\mathcal{A},\alpha,\tilde{\mathbf{p}}\left(\mathcal{A},\alpha\right)\right)<\rho_{\mathcal{A}_{j}}\right\} $
and $\mathcal{K}=\left\{ k:\:\tilde{p}_{k}\left(\mathcal{A},\alpha\right)>0\right\} $.
Note that we have $\tilde{\lambda}_{j}=0,\:\forall j\in\mathcal{J}$
and $\tilde{\nu}_{k}=0,\:\forall k\in\mathcal{K}$ according to the
KKT conditions. It can be verified that $\frac{\partial\tilde{\lambda}_{j}}{\partial\alpha}=0,\:\forall j\in\mathcal{J}$
and $\frac{\partial\tilde{\nu}_{k}}{\partial\alpha}=0,\:\forall k\in\mathcal{K}.$
Therefore, we can delete these $\left|\mathcal{J}\right|+\left|\mathcal{K}\right|$
variables and the corresponding linear equations whose index $i$
satisfies $i-K\in\mathcal{J}$ or $i-S-K\in\mathcal{K}$. The remaining
$2K+S-\left|\mathcal{J}\right|-\left|\mathcal{K}\right|$ variables
can be determined by the remaining linear equations. After obtaining
$\frac{\partial\tilde{\mathbf{p}}\left(\mathcal{A},\alpha\right)}{\partial\alpha}$,
the derivative $\frac{\partial\hat{\mathcal{I}}\left(\mathcal{A},\alpha\right)}{\partial\alpha}$
can be calculated using (\ref{eq:Idruo}).

To complete the calculation of $\frac{\partial\hat{\mathcal{I}}\left(\mathcal{A},\alpha\right)}{\partial\alpha}$,
we still need to obtain $\frac{\partial g_{kl}}{\partial\alpha},\:\forall k,l$,
and $\frac{\partial\mathbf{R}}{\partial\alpha}$. The following Lemma
are useful and can be proved by a direct calculation.
\begin{lem}
[Derivatives of the intermediate variables]\label{lem:Derivatives-ruo}For
the intermediate variables $\boldsymbol{\theta}_{k}$, $\mathbf{v}$,
$\mathbf{\Delta}$ and $\mathbf{C}$ defined in Lemma \ref{lem:AsyWS}
and Lemma \ref{lem:AsyPm}, the partial derivatives of them with respective
to $\alpha$ are given below.
\end{lem}
\begin{equation}
\frac{\partial\boldsymbol{\theta}_{k}}{\partial\alpha}=\left(\mathbf{I}_{K}-\mathbf{D}\right)^{-1}\left(\frac{\partial\mathbf{d}_{k}}{\partial\alpha}+\frac{\partial\mathbf{D}}{\partial\alpha}\boldsymbol{\theta}_{k}\right),\:\forall k,\label{eq:Thedruo}
\end{equation}
where $\frac{\partial\mathbf{d}_{k}}{\partial\alpha}$ is given by
\[
\frac{\partial d_{kl}}{\partial\alpha}=-\frac{2}{S}\sum_{m\in\mathcal{A}}\left[\sigma_{km}^{2}\sigma_{lm}^{2}\left(1-\frac{1}{S}\sum_{i=1}^{K}\frac{\sigma_{im}^{2}\varphi_{i}}{\left(1+\xi_{i}\right)^{2}}\right)/f_{m}^{3}\left(\boldsymbol{\xi}\right)\right],\:\forall l,
\]
and $\frac{\partial\mathbf{D}}{\partial\alpha}$ is given by
\[
\frac{\partial D_{ln}}{\partial\alpha}=\frac{1}{S^{2}}\sum_{m\in\mathcal{A}}\left[\frac{2\sigma_{lm}^{2}\sigma_{nm}^{2}}{\left(1+\xi_{n}\right)^{2}}\left[\frac{1}{S}\sum_{i=1}^{K}\frac{\sigma_{im}^{2}}{1+\xi_{i}}\left(\frac{\varphi_{i}}{1+\xi_{i}}-\frac{\varphi_{n}}{1+\xi_{n}}\right)-1-\frac{\alpha\varphi_{n}}{1+\xi_{n}}\right]/f_{m}^{3}\left(\boldsymbol{\xi}\right)\right].
\]
\begin{equation}
\frac{\partial C_{ln}}{\partial\alpha}=\frac{1}{S}\sum_{m\in\mathcal{A}}\left[\frac{1}{S}\sigma_{lm}^{2}\sigma_{nm}^{2}\frac{\partial v_{l}}{\partial\alpha}/\psi_{m}^{2}\left(\mathbf{v}\right)-\frac{2}{S}\sigma_{lm}^{2}\sigma_{nm}^{2}v_{l}\frac{1}{S}\sum_{i=1}^{K}\sigma_{im}^{2}\frac{\partial v_{i}}{\partial\alpha}/\psi_{m}^{3}\left(\mathbf{v}\right)\right],\:\forall l,n,\label{eq:Cdruo}
\end{equation}
\begin{equation}
\frac{\partial\triangle_{l}}{\partial\alpha}=-\frac{1}{S}\sum_{m\in\mathcal{A}}\left[\sigma_{lm}^{2}\frac{1}{S}\sum_{i=1}^{K}\sigma_{im}^{2}\frac{\partial v_{i}}{\partial\alpha}/\psi_{m}^{2}\left(\mathbf{v}\right)\right],\:\forall l,\label{eq:Sjdruo}
\end{equation}
\begin{equation}
\frac{\partial\mathbf{v}}{\partial\alpha}=-\left(\alpha\mathbf{I}_{K}+\mathbf{\Delta}-\mathbf{C}\right)^{-1}\mathbf{v},\label{eq:vduo}
\end{equation}

Using Lemma \ref{lem:Derivatives-ruo}, $\frac{\partial g_{kl}}{\partial\alpha},\:\forall k,l$,
and $\frac{\partial\mathbf{R}}{\partial\alpha}$ can be obtained by
a direct calculation as follows.
\begin{eqnarray*}
\frac{\partial g_{kk}}{\partial\alpha} & = & \frac{2\xi_{k}\frac{\partial\xi_{k}}{\partial\alpha}}{\left(1+\xi_{k}\right)^{3}},\:\forall k,\\
\frac{\partial g_{kl}}{\partial\alpha} & = & \frac{\frac{\partial\theta_{kl}}{\partial\alpha}}{S\left(1+\xi_{l}\right)^{2}\left(1+\xi_{k}\right)^{2}}-\frac{2\theta_{kl}\left[\frac{\partial\xi_{l}}{\partial\alpha}\left(1+\xi_{k}\right)+\frac{\partial\xi_{k}}{\partial\alpha}\left(1+\xi_{l}\right)\right]}{S\left(1+\xi_{l}\right)^{3}\left(1+\xi_{k}\right)^{3}},\:\forall k\neq l,
\end{eqnarray*}
where $\frac{\partial\xi_{k}}{\partial\alpha}=\varphi_{k},\:\forall k$
is defined in (\ref{eq:FaiDef}), $\boldsymbol{\theta}_{k}=\left[\theta_{k1},...,\theta_{kK}\right]^{T},\:\forall k$
is defined in (\ref{eq:ThetaLE}) and $\frac{\partial\boldsymbol{\theta}_{k}}{\partial\alpha}$
is given in (\ref{eq:Thedruo}). To calculate $\frac{\partial\mathbf{R}}{\partial\alpha}$,
we first obtain $\frac{\partial\hat{\mathbf{R}}}{\partial\alpha}$
as
\[
\frac{\partial\hat{R}_{kj}}{\partial\alpha}=-\frac{\sigma_{k\mathcal{A}_{j}}^{2}\alpha^{-1}}{S^{2}}\left[\alpha^{-1}/\psi_{\mathcal{A}_{j}}^{2}\left(\mathbf{v}\right)+\frac{2}{S}\sum_{i=1}^{K}\sigma_{i\mathcal{A}_{j}}^{2}\frac{\partial v_{i}}{\partial\alpha}/\psi_{\mathcal{A}_{j}}^{3}\left(\mathbf{v}\right)\right],\:\forall k,j.
\]
where $\frac{\partial v_{i}}{\partial\alpha},\:\forall i$ is given
in (\ref{eq:vduo}). Then we obtain $\frac{\partial\bar{\mathbf{R}}}{\partial\alpha}$
as 
\begin{eqnarray*}
\frac{\partial\bar{R}_{kk}}{\partial\alpha} & = & \frac{1}{S}\sum_{m\in\mathcal{A}}\left[\left(\sigma_{km}^{2}\frac{\partial v_{k}}{\partial\alpha}+\sigma_{km}^{2}\frac{1}{S}\sum_{i\neq k}^{K}\sigma_{im}^{2}\left(v_{k}\frac{\partial v_{i}}{\partial\alpha}+v_{i}\frac{\partial v_{k}}{\partial\alpha}\right)\right)/\psi_{m}^{2}\left(\mathbf{v}\right)\right.\\
 &  & \left.-\left(2\sigma_{km}^{2}v_{k}\left(1+\frac{1}{S}\sum_{i\neq k}^{K}\sigma_{im}^{2}v_{i}\right)\frac{1}{S}\sum_{i=1}^{K}\sigma_{im}^{2}\frac{\partial v_{i}}{\partial\alpha}\right)/\psi_{m}^{3}\left(\mathbf{v}\right)\right],\forall k,\\
\frac{\partial\bar{R}_{kl}}{\partial\alpha} & = & -\frac{1}{S}\sum_{m\in\mathcal{A}}\left[\sigma_{km}^{2}\sigma_{lm}^{2}\frac{1}{S}\left(v_{l}\frac{\partial v_{k}}{\partial\alpha}+v_{k}\frac{\partial v_{l}}{\partial\alpha}\right)/\psi_{m}^{2}\left(\mathbf{v}\right)-\frac{2}{S}\sigma_{km}^{2}\sigma_{lm}^{2}v_{k}v_{l}\frac{1}{S}\sum_{i=1}^{K}\sigma_{im}^{2}\frac{\partial v_{i}}{\partial\alpha}/\psi_{m}^{3}\left(\mathbf{v}\right)\right],\forall k\neq l.
\end{eqnarray*}
Finally, $\frac{\partial\mathbf{R}}{\partial\alpha}=\left[\begin{array}{cc}
\mathbf{I}_{S}, & \mathbf{1}\end{array}\right]^{T}\frac{\partial\tilde{\mathbf{R}}}{\partial\alpha}$ and $\frac{\partial\tilde{\mathbf{R}}}{\partial\alpha}$ is given
by
\begin{eqnarray*}
\frac{\partial\tilde{\mathbf{R}}}{\partial\alpha} & = & \left(\frac{\partial\hat{\mathbf{R}}}{\partial\alpha}\right)^{T}\left(\mathbf{V}-\left(\alpha\mathbf{I}_{K}+\mathbf{\Delta}-\mathbf{C}\right)^{-1}\bar{\mathbf{R}}\right)+\hat{\mathbf{R}}^{T}\left(\frac{\partial\mathbf{V}}{\partial\alpha}-\left(\alpha\mathbf{I}_{K}+\mathbf{\Delta}-\mathbf{C}\right)^{-1}\frac{\partial\bar{\mathbf{R}}}{\partial\alpha}\right)\\
 &  & +\hat{\mathbf{R}}^{T}\left(\mathbf{I}_{K}+\frac{\partial\mathbf{\Delta}}{\partial\alpha}-\frac{\partial\mathbf{\mathbf{C}}}{\partial\alpha}\right)\left(\alpha\mathbf{I}_{K}+\mathbf{\Delta}-\mathbf{C}\right)^{-2}\bar{\mathbf{R}},
\end{eqnarray*}
where $\frac{\partial\mathbf{\mathbf{C}}}{\partial\alpha}$ and $\frac{\partial\mathbf{\Delta}}{\partial\alpha}$
are given in (\ref{eq:Cdruo}) and (\ref{eq:Sjdruo}) respectively,
and $\frac{\partial\mathbf{V}}{\partial\alpha}=\textrm{diag}\left(\frac{\partial v_{1}}{\partial\alpha},...,\frac{\partial v_{K}}{\partial\alpha}\right)$.

\subsection{Proof of Theorem \ref{thm:OptruoCA}\label{sub:Proof-of-TheoremOptruoCA}}

When $P_{T}$ is large enough, all users will be allocated with non-zero
power. In this case, the SINR of user $k$ under power allocation
in (\ref{eq:pkCA}) is given by
\begin{eqnarray}
 &  & \hat{\gamma}_{k}\left(\alpha,\mathbf{p}^{*}\left(\mathcal{A},\alpha\right)\right)\label{eq:SinrCA}\\
 & = & \frac{Sw_{k}\sigma_{k1}^{2}\left(2\alpha uP_{T}+\left(\beta-1\right)P_{T}+\frac{1}{S}\sum_{l=1}^{K}\frac{1}{\sigma_{l1}^{2}}\right)}{\left(\sum_{l=1}^{K}w_{l}\right)\left(\frac{\sigma_{k1}^{2}P_{T}}{\left(1+\sigma_{k1}^{2}u\right)^{2}}+1\right)}-1,\nonumber 
\end{eqnarray}
and $\bar{\mathcal{I}}\left(\mathcal{A},\alpha,\mathbf{p}^{*}\left(\mathcal{A},\alpha\right)\right)=\sum_{k=1}^{K}w_{k}\textrm{log}\left(1+\hat{\gamma}_{k}\left(\alpha,\mathbf{p}^{*}\left(\mathcal{A},\alpha\right)\right)\right)$.
For any $k$, it can be shown that the solution $\tilde{\alpha}_{k}$
of $\frac{\partial}{\partial\alpha}\hat{\gamma}_{k}\left(\alpha,\mathbf{p}^{*}\left(\mathcal{A},\alpha\right)\right)=0$
must satisfy $\tilde{\alpha}_{k}=O\left(\frac{1}{P_{T}}\right)$.
Since the optimal regularization factor $\alpha^{*}$ must satisfy
$\underset{k}{\textrm{min}}\tilde{\alpha}_{k}\leq\alpha^{*}\leq\underset{k}{\textrm{max}}\tilde{\alpha}_{k}$,
we have $\alpha^{*}=O\left(\frac{1}{P_{T}}\right)$. To prove the
second result, it can be verified that $\frac{\partial^{2}}{\partial^{2}\alpha}\hat{\gamma}_{k}\left(\alpha,\mathbf{p}^{*}\left(\mathcal{A},\alpha\right)\right)<0$
and thus $\hat{\gamma}_{k}\left(\alpha,\mathbf{p}^{*}\left(\mathcal{A},\alpha\right)\right)$
is concave when $\alpha$ is small enough. Since $\bar{\mathcal{I}}\left(\mathcal{A},\alpha,\mathbf{p}^{*}\left(\mathcal{A},\alpha\right)\right)$
is a concave increasing function of $\hat{\gamma}_{k}\left(\alpha,\mathbf{p}^{*}\left(\mathcal{A},\alpha\right)\right)$,
$\bar{\mathcal{I}}\left(\mathcal{A},\alpha,\mathbf{p}^{*}\left(\mathcal{A},\alpha\right)\right)$
must be a concave function of $\alpha$ \cite{Boyd_04Book_Convex_optimization}.

\subsection{Proof of Theorem \ref{thm:Asymptotic-Decoupling}\label{sub:Sketch-ThmDecoupling}}

We first derive a lower bound for the probability that the minimum
distance $r_{\textrm{min}}$ between any two users is larger than
a certain value $r_{0}$: $\textrm{Pr}\left(r_{\textrm{min}}\geq r_{0}\right)$.
Let $d_{kl}^{u}$ denote the distance between user $k$ and user $l$.
We have 
\begin{eqnarray*}
\textrm{Pr}\left(r_{\textrm{min}}\geq r_{0}\right) & = & 1-\textrm{Pr}\left(\underset{l\neq k}{\textrm{min}}\: d_{kl}^{u}\leq r_{0},\:\exists k\in\left\{ 1,...,K\right\} \right)\\
 & \geq & 1-\sum_{k=1}^{K}\textrm{Pr}\left(\underset{l\neq k}{\textrm{min}}\: d_{kl}^{u}\leq r_{0}\right)\\
 & = & 1-K\textrm{Pr}\left(\underset{l\neq1}{\textrm{min}}\: d_{1l}^{u}\leq r_{0}\right)\\
 & = & 1-K\left[1-\left(\textrm{Pr}\left(d_{12}^{u}\geq r_{0}\right)\right)^{K-1}\right]\\
 & \geq & 1-K\left[1-\left(1-\frac{\pi r_{0}^{2}}{R_{c}^{2}}\right)^{K-1}\right],
\end{eqnarray*}
where the second inequality follows from the union bound and the last
inequality holds because $\textrm{Pr}\left(d_{12}^{u}\geq r_{0}\right)$
$\geq1-\frac{\pi r_{0}^{2}}{R_{c}^{2}}$. 

Then we use the path loss model to transfer the probability $\textrm{Pr}\left(r_{\textrm{min}}\geq r_{0}\right)$
to the probability $\textrm{Pr}\left(\eta>\eta_{0}\right)$ in (\ref{eq:ProDCgain}).
Note that for any $k$, we have $\underset{m}{\textrm{min}}\: r_{km}\leq\frac{\sqrt{2}R_{c}}{2\sqrt{M}}$,
and $\underset{l}{\textrm{max}}\: r_{k\tilde{m}_{l}}\geq r_{\textrm{min}}-\frac{\sqrt{2}R_{c}}{2\sqrt{M}}$.
Hence $\eta\geq\left(r_{\textrm{min}}/\left(\frac{\sqrt{2}R_{c}}{2\sqrt{M}}\right)-1\right)^{\zeta}$
and
\begin{eqnarray*}
\textrm{Pr}\left(\eta>\eta_{0}\right) & \geq & \textrm{Pr}\left(\left(r_{\textrm{min}}/\left(\frac{\sqrt{2}R_{c}}{2\sqrt{M}}\right)-1\right)^{\zeta}>\eta_{0}\right)\\
 & = & \textrm{Pr}\left(r_{\textrm{min}}>\frac{\sqrt{2}R_{c}}{2\sqrt{M}}\left(\eta_{0}^{1/\zeta}+1\right)\right)\\
 & \geq & 1-K\left[1-\left(1-\frac{\pi\left(\eta_{0}^{1/\zeta}+1\right)^{2}}{2M}\right)^{K-1}\right].
\end{eqnarray*}

Finally, we prove the capacity scaling by deriving an upper and a
lower bound for the sum-rate. The following lemma follows directly
from Definition \ref{def:VLDAS}. 
\begin{lem}
\label{lem:gdUp}For any $\epsilon>0$, as $M\rightarrow\infty$ with
$K,S$ fixed, we have
\[
\textrm{Pr}\left(\underset{k,m}{\textrm{min}}\: r_{km}\leq M^{-\frac{1}{2}-\epsilon}\right)=\frac{\pi M^{-2\epsilon}}{R_{c}^{2}}\rightarrow0,
\]
and thus
\[
\textrm{Pr}\left(\bar{g}_{k}^{d}>G_{0}M^{\frac{\zeta}{2}+\epsilon}\right)\rightarrow0.
\]

\end{lem}

Let $P_{T}^{U}=\underset{m\in\mathcal{A}}{\textrm{max}}\rho_{m}$
and let $X_{S}$ be a random variable with $\chi^{2}\left(2S\right)$
distribution. Assuming that each user is served by $S$ antennas without
mutual interference, we obtain an upper bound for average sum-rate
as follows:
\begin{eqnarray}
C_{s} & \leq & K\textrm{E}\left[\textrm{log}\left(1+P_{T}^{U}\bar{g}_{k}^{d}X_{S}/2\right)\right]\nonumber \\
 & \leq & K\textrm{log}\left(1+P_{T}^{U}\bar{g}_{k}^{d}\textrm{E}\left[X_{S}\right]/2\right).\label{eq:CwUp}
\end{eqnarray}
Combining (\ref{eq:CwUp}) with Lemma \ref{lem:gdUp}, we prove that
$C_{s}\overset{a.s}{\leq}O\left(K\left(\frac{\zeta}{2}+\epsilon\right)\textrm{log}M\right)$
as $M\rightarrow\infty$ with $K,S$ fixed.

Furthermore, it follows from the lower bound provided in Appendix
\ref{sub:Sketch-OptS3} that $C_{s}\geq O\left(K\left(\frac{\zeta}{2}-\epsilon\right)\textrm{log}M\right)$.
This completes the proof of Theorem \ref{thm:Asymptotic-Decoupling}.

\subsection{Proof of Corollary \ref{cor:Asymptotic-Optimality-ofS3}\label{sub:Sketch-OptS3}}

Due to (\ref{eq:ProDCgain}) in Theorem \ref{thm:Asymptotic-Decoupling},
the step 1 in Algorithm S2 will almost surely select a set of antennas
$\mathcal{A},\:\left|\mathcal{A}\right|=K$ such that each user has
strong direct-link with one of the selected $K$ antennas and weak
cross-links with other selected antennas for large $M/K$. Assume
that each selected antenna only serves the nearest user, and assume
equal power allocation for each user, i.e., the transmit power for
each user is $P^{L}=\underset{m\in\mathcal{A}}{\textrm{min}}\frac{\rho_{m}}{S}$.
Let $X_{m}$ denote a random variable with $\chi^{2}\left(2m\right)$
distribution. Let $\eta_{0}=M^{\frac{\zeta}{2}-\epsilon_{1}}$ in
(\ref{eq:ProDCgain}). Then using (\ref{eq:ProDCgain}) and the fact
that $\bar{g}_{k}^{d}\geq G_{0}\left(\frac{\sqrt{2}R_{c}}{2\sqrt{M}}\right)^{-\zeta}$,
we can show that as $M\rightarrow\infty$ with $K,S$ fixed, the average
sum-rate $\mathcal{I}_{\mathcal{A}}$ is almost surely lower bounded
by 
\begin{equation}
\mathcal{I}_{\mathcal{A}}\overset{a.s}{\geq}K\textrm{E}\left[\textrm{log}\left(1+\frac{P^{L}G_{0}\left(\frac{\sqrt{2}R_{c}}{2\sqrt{M}}\right)^{-\zeta}X_{1}/2}{1+P^{L}M^{-\frac{\zeta}{2}+\epsilon_{1}}G_{0}\left(\frac{\sqrt{2}R_{c}}{2\sqrt{M}}\right)^{-\zeta}\frac{X_{K-1}}{2}}\right)\right],\label{eq:Ilowb1}
\end{equation}
where $X_{1}$ and $X_{K-1}$ are independent. Choose $B_{1},B_{2}>0$
such that $\textrm{Pr}\left(X_{1}\geq B_{1}\right)\textrm{Pr}\left(X_{K-1}\leq B_{2}\right)\geq1-\epsilon_{2}$.
Then as $M\rightarrow\infty$ with $K,S$ fixed, it follows from (\ref{eq:Ilowb1})
that {\small{
\begin{eqnarray*}
\mathcal{I}_{\mathcal{A}} & \overset{a.s}{\geq} & K\left(1-\epsilon_{2}\right)\textrm{log}\left(1+\frac{P^{L}G_{0}\left(\frac{\sqrt{2}R_{c}}{2\sqrt{M}}\right)^{-\zeta}B_{1}/2}{1+P^{L}M^{-\frac{\zeta}{2}+\epsilon_{1}}G_{0}\left(\frac{\sqrt{2}R_{c}}{2\sqrt{M}}\right)^{-\zeta}\frac{B_{2}}{2}}\right)\\
 & = & O\left(K\left(1-\epsilon_{2}\right)\left(\frac{\zeta}{2}-\epsilon_{1}\right)\textrm{log}M\right).
\end{eqnarray*}
}}Choose $\epsilon_{1},\epsilon_{2}$ such that $\epsilon_{1}+\frac{\zeta}{2}\epsilon_{2}-\epsilon_{1}\epsilon_{2}=\epsilon$.
Then we have $\mathcal{I}_{\mathcal{A}}\overset{a.s}{\geq}O\left(K\left(\frac{\zeta}{2}-\epsilon\right)\textrm{log}M\right)$
as $M\rightarrow\infty$ with $K,S$ fixed. The rest of the steps
in S2 only increase the sum-rate by a constant. This completes the
proof.%

\end{document}